\documentclass[aip,amsmath,amsfonts,amssymb,reprint]{revtex4-1}
\usepackage{physics}
\usepackage{mathtools}
\usepackage{graphicx}
\usepackage{epsf}
\usepackage{braket}
\usepackage{epstopdf}
\usepackage{setspace}
\usepackage{array}
\usepackage{float}
\usepackage{hyperref}
\usepackage{color}
\usepackage{bm}
\usepackage{amsmath}
\usepackage{mathtools}
\usepackage{bigints}
\usepackage{nicematrix}
\usepackage{xcolor}
\usepackage{lipsum}
\usepackage{textcomp}
\usepackage{ragged2e}
\usepackage{fancybox}
\usepackage{enumitem}
\usepackage{dcolumn}
\usepackage{subcaption}
\begin{document}
	
	\title{\textcolor{black}{Band structure evolution from kagome to Lieb under periodic driving field}}
	
	\author{Gulshan Kumar}
	\affiliation{Department of Physics,
		Indian Institute of Technology Patna, Bihta, Bihar, 801106, India}
	\author{Shashikant Kumar}
	\affiliation{Department of Physics,
		Indian Institute of Technology Patna, Bihta, Bihar, 801106, India}
	\author{Prakash Parida}\email{pparida@iitp.ac.in}
	\affiliation{Department of Physics,
		Indian Institute of Technology Patna, Bihta, Bihar, 801106, India}
	%\date{\today}
	\begin{abstract}
    We theoretically investigate the light-induced transition of the kagome quasienergy spectrum to the Lieb like band structure under periodic driving fields. A generalized framework for the renormalized hopping potential is derived, applicable to any two-dimensional lattice with arbitrary field polarizations. By applying this framework to a kagome lattice driven by linearly polarized light in off-resonant condition, we demonstrate the ability to tune the hopping strength along specific bonds to zero. This tuning induces the merging of Dirac points at high-symmetry points in the Brillouin zone, governed by the field parameters. At specific parameter values, this merging facilitates a transition from the kagome quasienergy spectrum to the Lieb band structure with reduced bandwidth. Our results highlight the critical role of controlled electron hopping in driving this electronic transition, offering valuable insights into the manipulation of electronic properties in periodically driven systems.		
	\end{abstract}
	
	%\section*{keywords}{Suggested keywords}%Use showkeys class option if keyword
	%display desired
	\maketitle
	Flat-band systems have gained significant attention in theoretical and experimental research \cite{the_flat1,the_flat2,exp_flat1,exp_flat2} due to their high density of states, which promotes strong electron correlations and enables unique physical phenomena \cite{photonics_flat1,photonics_flat2,electronic_flat1,electronic_flat2,ferro_flat1,ferro_flat2}. Among these quantum systems, two-dimensional (2D) flat-band systems are the most studied, as their physical properties can be tuned using external perturbations \cite{exp_flat2,sys_flat1,sys_flat2,sys_flat3,sys_flat4,sys_flat5}. Mechanical strain \cite{strain_1,strain_2,strain_3,strain_4,strain_5}, in particular, allows precise manipulation of flat band physics by altering hopping energy through inducig structural anisotropy \cite{anisotropy_1,anisotropy_2}.	
	
	The effects of strain on flat-band systems, particularly kagome and Lieb lattices, have been extensively studied \cite{lieb_strain1, flat_strain1, strain_kagome1, flat_kagome_lieb2, flat_kagome_lieb1}. Jiang $et \, al. $ showed that diagonal strain can transform Lieb lattices into kagome lattices due to their shared unit cell structures \cite{flat_kagome_lieb2}. However, Lima $et \, al. $ demonstrated that uniaxial, biaxial, and shear strain distort flat and dispersive bands, introducing anisotropy without opening energy gaps. Additionally, in Lieb lattices, strain splits triply degenerate Dirac points into two doubly degenerate points \cite{strain_kagome_lieb1}. Similarly, Xu $et \, al. $ revealed that uniaxial strain shifts Dirac points in kagome lattices without merging them, unlike in graphene \cite{strain_kagome2}. However, a key limitation of strain engineering is that it must remain within elastic limits to avoid damaging materials. Moreover, it cannot entirely suppress hopping along specific bonds without compromising lattice integrity \cite{strain_damage}. This highlights the challenges in tuning flat-band systems using strain while maintaining the structural and electronic properties of the material.
	
	Periodic driving fields overcome these limitations by inducing anisotropy in hopping strength without structural damage \cite{merging}. Studies on kagome lattices under circularly polarized fields, such as those by He et al., observed quasienergy gaps bridged by chiral edge states \cite{flat_floquet_kagome2}. Du et al. found that circular polarization creates gaps via two-photon processes, while linear polarization splits quadratic band touchings into Dirac points \cite{flat_floquet1}. Despite these advances, the merging of Dirac points in kagome lattices under periodic driving remains unexplored. 
	\begin{figure*}[ht]
		\centering
		\begin{subfigure}[b]{0.22\textwidth}
			%   \centering
			\subcaption{}
			\includegraphics[width=4.2cm,height=3cm]{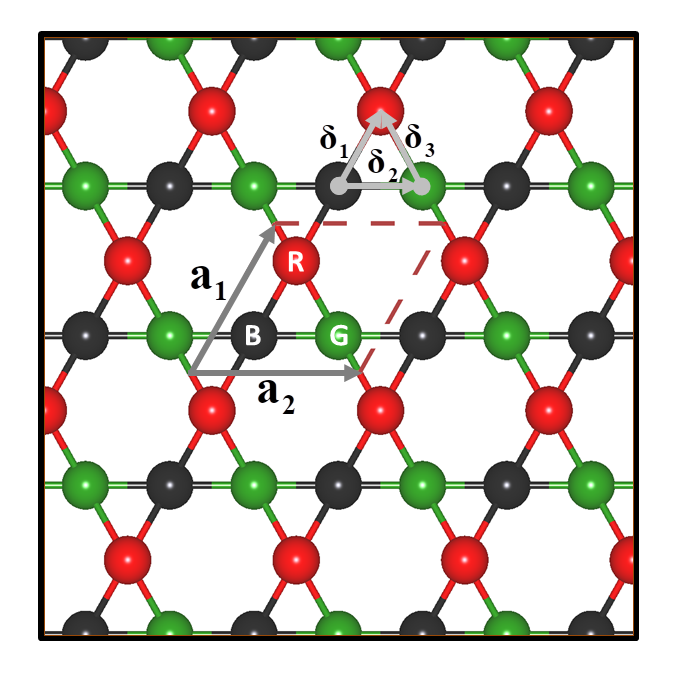}
			\label{fig:kagome_lattice}
		\end{subfigure}
		\begin{subfigure}[b]{0.2\textwidth}
			%      \centering
			\subcaption{}
			\includegraphics[width=3.5cm,height=3cm]{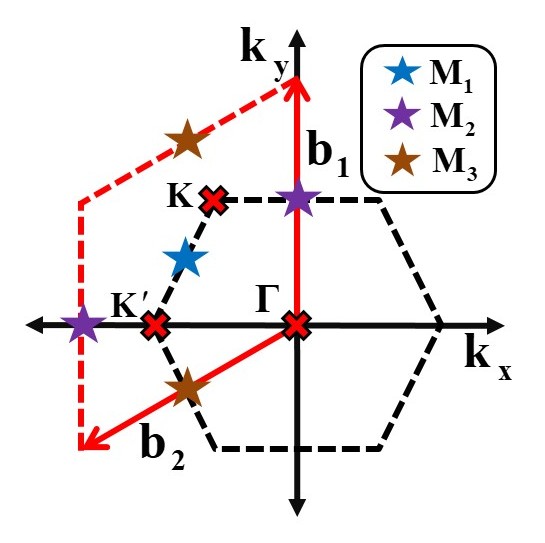}
			\label{fig:kagome_hsp}
		\end{subfigure}
		\begin{subfigure}[b]{0.24\textwidth}
			%      \centering
			\subcaption{}
			\includegraphics[width=4.25cm,height=3cm]{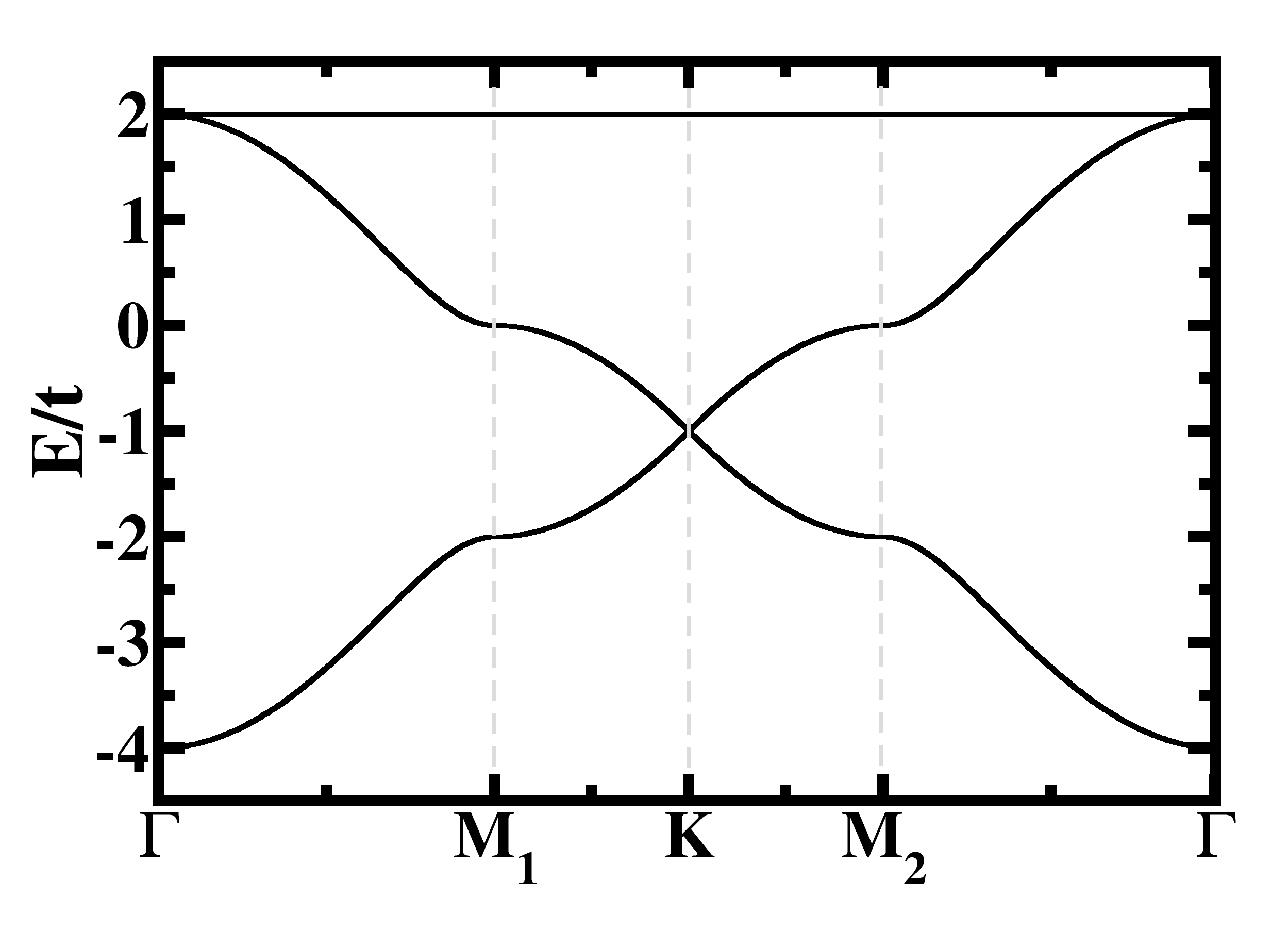}
			\label{fig:kagome_band_normal}
		\end{subfigure} 	
		\begin{subfigure}[b]{0.24\textwidth}
			%      \centering
			\subcaption{}
			\includegraphics[width=4.25cm,height=3cm]{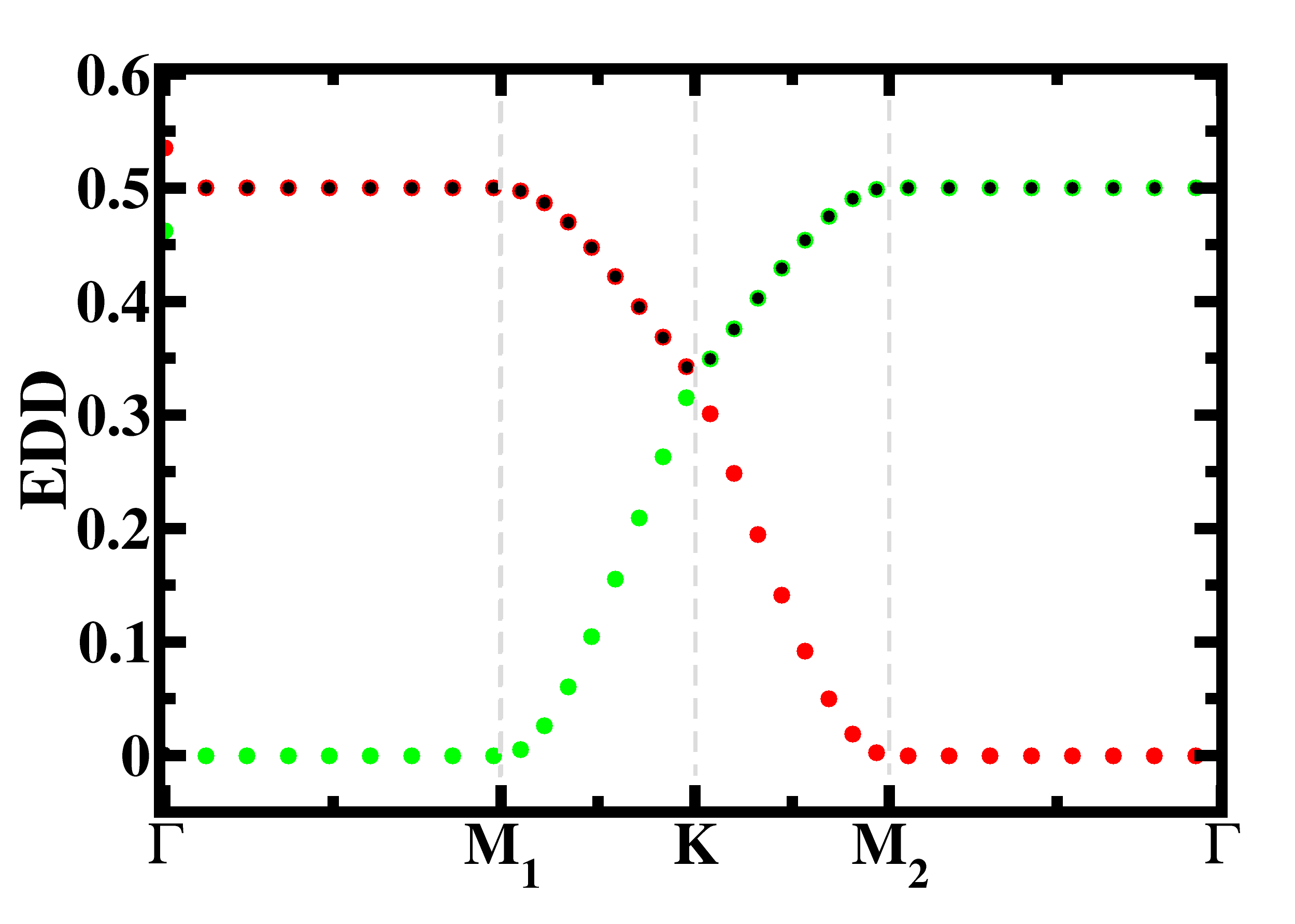}
			\label{fig:kagome_edd_flatband_normal}
		\end{subfigure}
		\begin{subfigure}[b]{0.22\textwidth}
			%   \centering
			\subcaption{}
			\includegraphics[width=4.2cm,height=3cm]{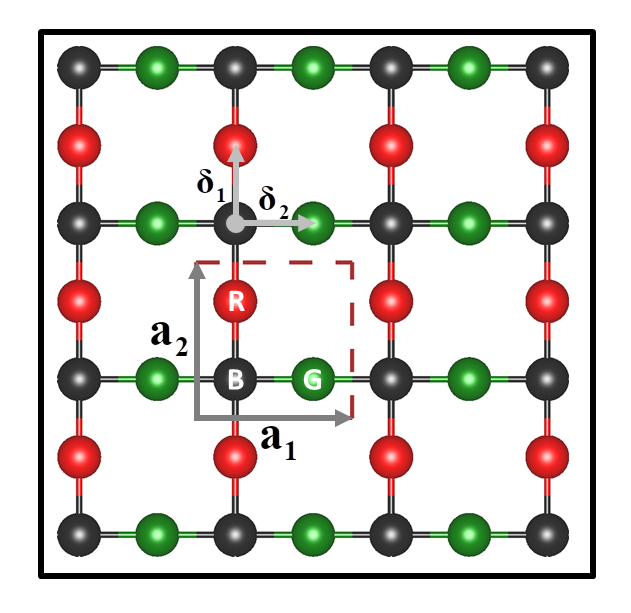}
			\label{fig:lieb_lattice}
		\end{subfigure}
		\begin{subfigure}[b]{0.2\textwidth}
			%   \centering
			\subcaption{}
			\includegraphics[width=3.5cm,height=3cm]{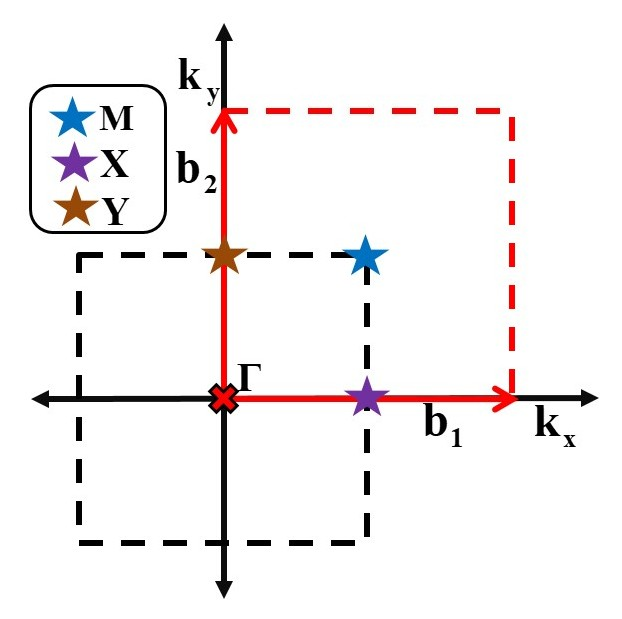}
			\label{fig:lieb_hsp}
		\end{subfigure}
		\begin{subfigure}[b]{0.24\textwidth}
			%   \centering
			\subcaption{}
			\includegraphics[width=4.25cm,height=3cm]{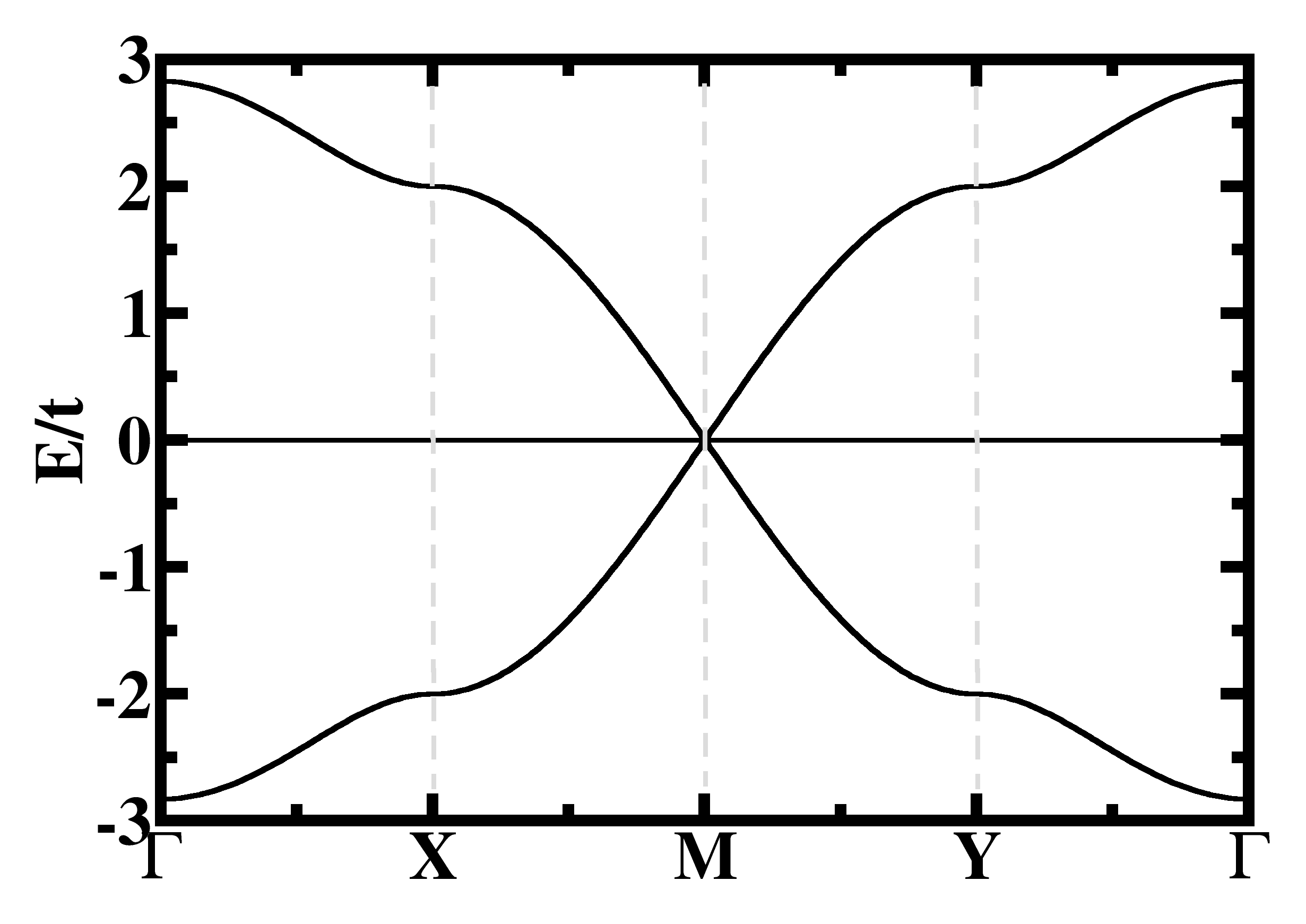}
			\label{fig:lieb_band_normal}
		\end{subfigure}
		\begin{subfigure}[b]{0.24\textwidth}
			%      \centering
			\subcaption{}
			\includegraphics[width=4.25cm,height=3cm]{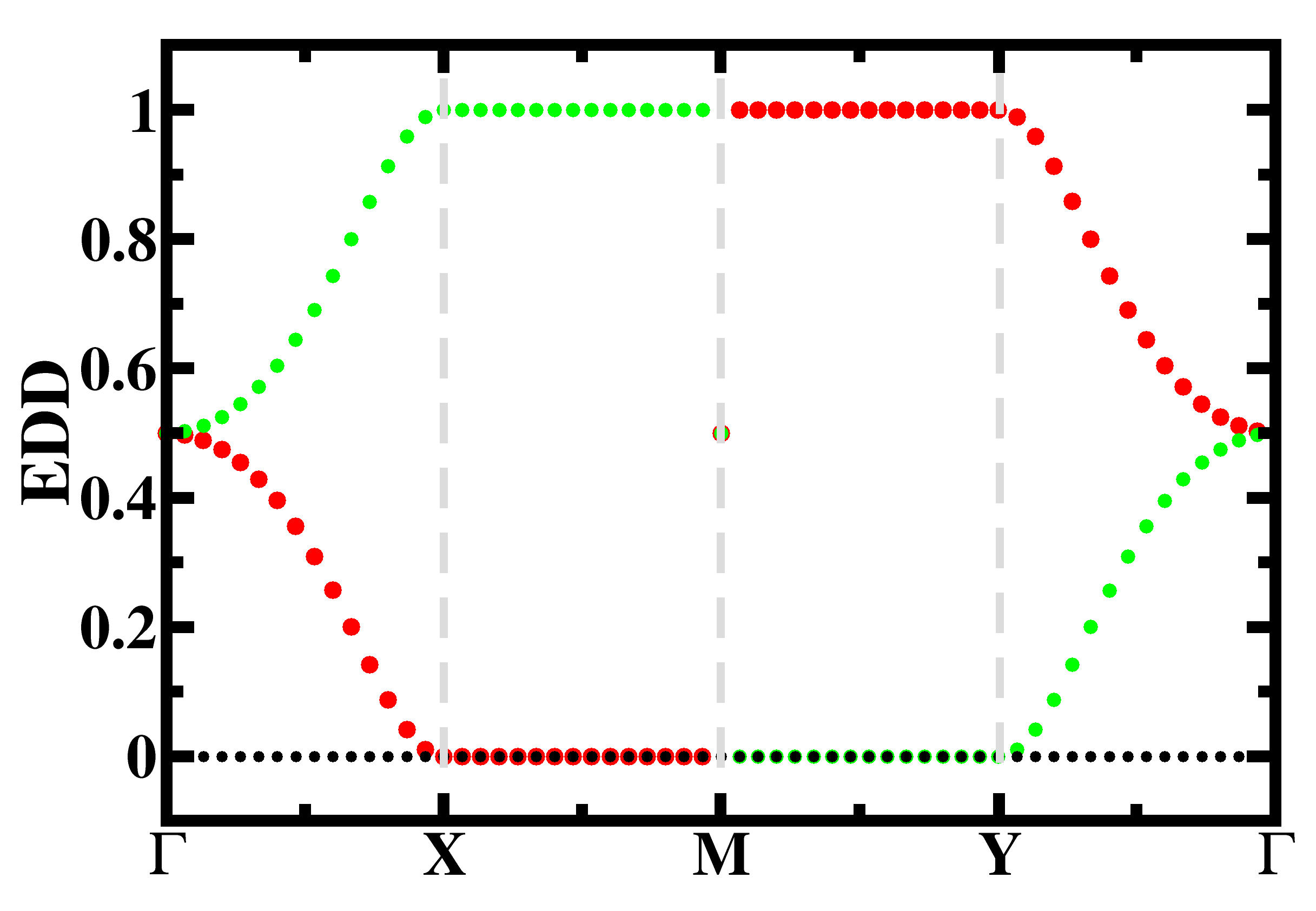}
			\label{fig:lieb_edd_flatband_normal}
		\end{subfigure}
		\caption{\label{fig:kagome_lieb_without_light} \justifying The 2D (a) kagome and (e) Lieb lattices are shown with their respective Brillouin zones in (b) and (f). Panels (c) and (g) display their tight-binding band structures, while (d) and (h) show flat band EDDs, with colors indicating atomic contributions. Energy E depicted to the undriven band.}
	\end{figure*}
	
   In this letter, we first presents a generalized formulation for renormalized hopping strength in 2D systems under periodic driving fields with arbitrary polarization, offering insights into light-dressed systems and enabling the construction of Floquet-Bloch Hamiltonians governing quasienergy spectra for any 2D structure. Using Floquet-Bloch formalism, we demonstrate that, under linearly polarised light (LPL), Dirac cone merge in the kagome lattice. Intrestingly, this merging finally leading to the evolution of kagome quasienergy structure into a Lieb band spectrum. The Floquet quasienergy spectrum may be observed experimentally using time- and angle-resolved photoemission spectroscopy (tr-ARPES) like other 2D materials \cite{tr_ARPES1,tr_ARPES2,tr_ARPES3,tr_ARPES4,tr_ARPES5}.
	
	We consider a 2D system defined by generalized lattice vectors (LVs) $\vb{a}_i = (\mu_{x,i} a_{x,i}, \mu_{y,i} a_{y,i})$ and nearest neighbor vectors (NNVs) $\vb{\delta}_j = (\nu_{x,j} \delta_{x,j}, \nu_{y,j} \delta_{y,j})$, where $a_{x,i}, a_{y,i}, \delta_{x,j}, \delta_{y,j}$ are magnitudes of the respective vector components, and $\mu, \nu = \pm 1$ determine their signs. When interacting with an electromagnetic field $\vb{A}(\tau) = (A_x \cos(\omega \tau), A_y \cos(\omega \tau + \phi))$, the hopping strength of electron from $i^{th}$ to $j^{th}$ site ($ t_{i,j} $) is modified using the Peierls substitution: $t_{i,j} \rightarrow t_{i,j} e^{i \int_{i}^{j} \vb{A}(\tau) \cdot d\vb{r}}$. This introduces time-translational invariance, allowing the Hamiltonian $\vb{H}(\vb{r},\tau)$ to be periodic in both space and time and treated using the Floquet-Bloch formalism. A detailed explanation of Floquet-Bloch formalism is provided in Sec. 1 of the supplementary materials (SM). The time-dependent Hamiltonian can be further expressed in the reciprocal space as a time-dependent Bloch Hamiltonian ($\vb{H}_{\vb{k}}(\tau)$) using a Fourier transformation. This Hamiltonian can then be substituted into Eq. S(4) (SM), which, upon solving, yields the renormalized hopping term that is independent of time. In this work we have derived a generalized expression for the renormalized hopping integral, which can be applicable to any 2D system along any NN vectors, regardless of the field polarization. This is given by:
	\begin{equation}\label{general_hopping}
		\begin{split}
			&t_{q,l}^{F} = t_{l} e^{i q\left(\nu_{x,l}\frac{ \pi}{2} +\tan^{-1}\left(\frac{-\nu_{x,l}\nu_{y,l}\beta_{y,l}\sin{(\phi)}}{\beta_{x,l}+\nu_{x,l}\nu_{y,l}\beta_{y,l}\cos{(\phi)}}\right)\right)}\\ 
			& \quad \times J_{q}\left( \sqrt{\beta_{x,l}^2 + \beta_{y,l}^2 + 2\nu_{x,l}\nu_{y,l}\beta_{x,l}\beta_{y,l}\cos{(\phi)}}\right).	
		\end{split}
	\end{equation}
	
	Here, $t_l$ represents the hopping strength of undriven lattice along $\delta_l$ bonds, while $J_q$ denotes the Bessel function of the $q$-th order, with $\beta_{x,l} = \delta_{x,l} A_x$ and $\beta_{y,l} = \delta_{y,l} A_y$. Natural units such as $\hbar$, $c$, and $e$ are set to $1$ for furthur calculations in this paper. 
	
	Using Eq. \ref{general_hopping}, each elements of the infinite dimension Floquet-Bloch Hamiltonian (FBH) matrix (Eq. S6 in SM) for any 2D system can be constructed and diagonalized to obtain quasienergy spectra. An important point to note is that when an NNV aligns with a coordinate axis, the orthogonal axis coefficient becomes zero, and we conventionally assign $\nu = +1$. This general formula is validated by reproducing results for graphene \cite{predicting, merging,cpb}. Furthermore, we use Eq. (\ref{general_hopping}) to derive modified hopping integrals for the kagome lattice and utilize it to generate quasienergy spectra of the system.

        \begin{figure}[ht]
        	\centering
        	\begin{subfigure}[b]{0.15\textwidth}
        		%   \centering
        		\subcaption{}
        		\includegraphics[width=2.8cm,height=2.2cm]{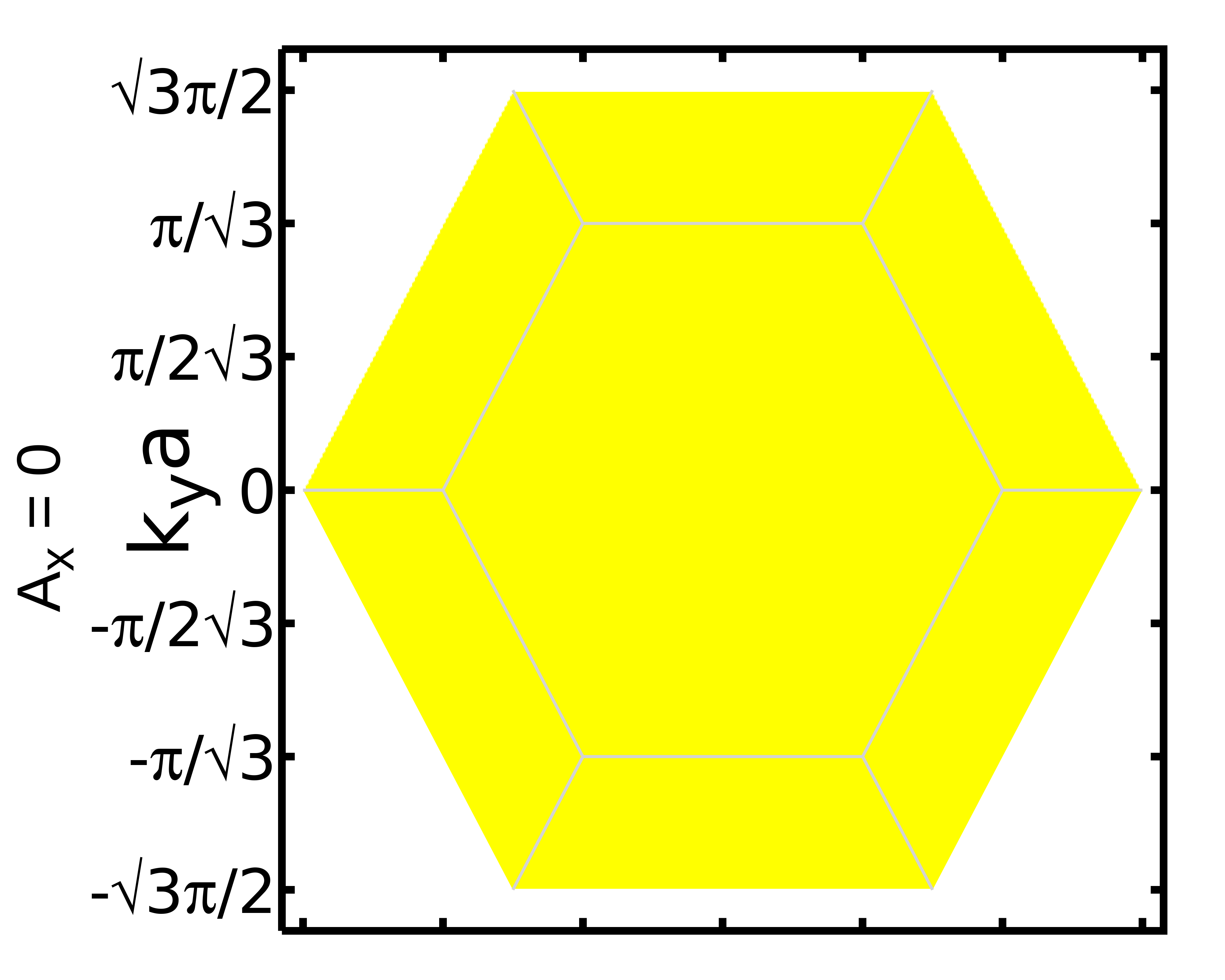}
        		%		\label{fig:kagome_lattice_delta1}
        	\end{subfigure}
        	\begin{subfigure}[b]{0.15\textwidth}
        		%   \centering
        		\subcaption{}
        		\includegraphics[width=2.5cm,height=2.2cm]{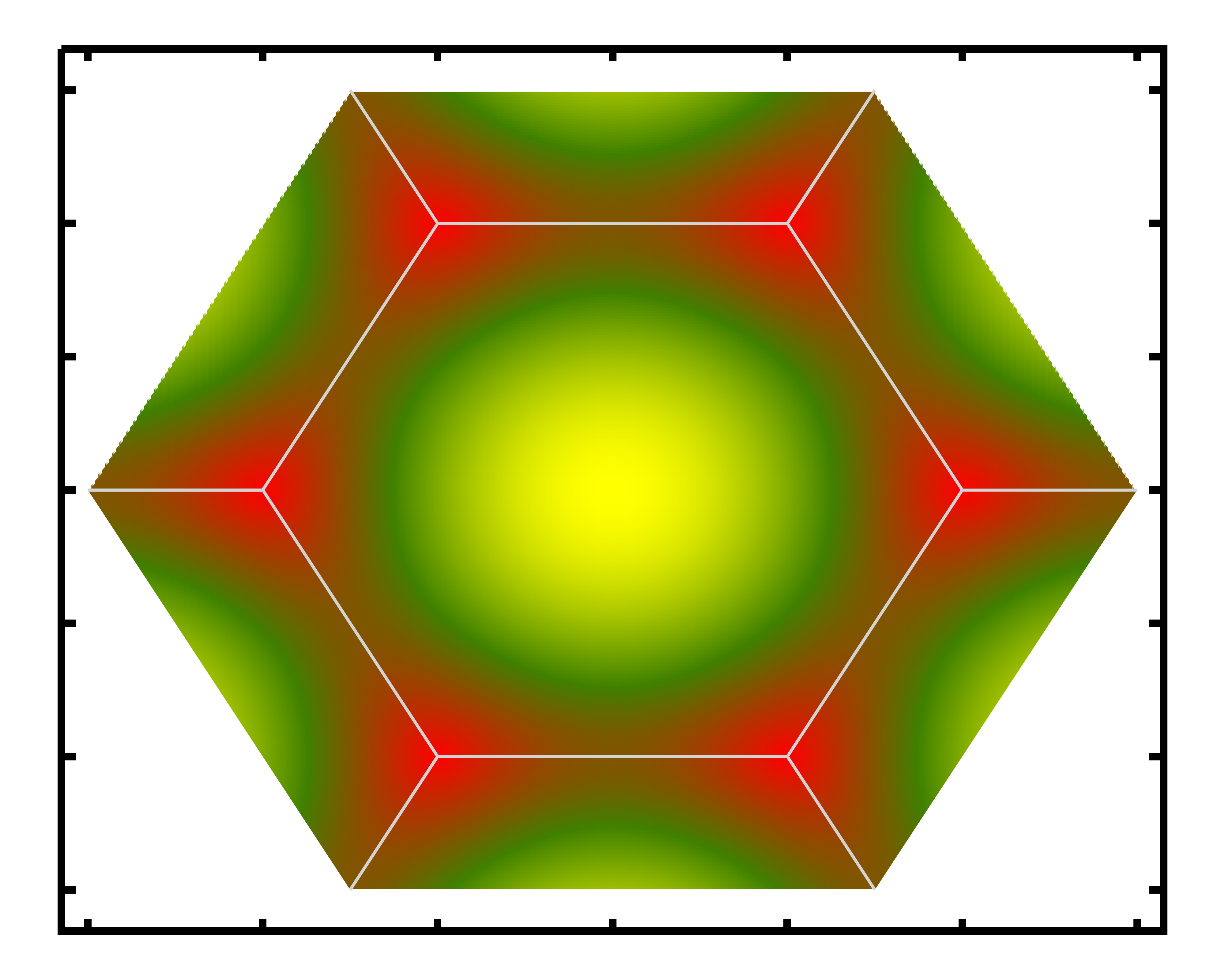}
        		%		\label{fig:kagome_lattice_delta2}
        	\end{subfigure}
        	\begin{subfigure}[b]{0.15\textwidth}
        		%   \centering
        		\subcaption{}
        		\includegraphics[width=2.8cm,height=2.2cm]{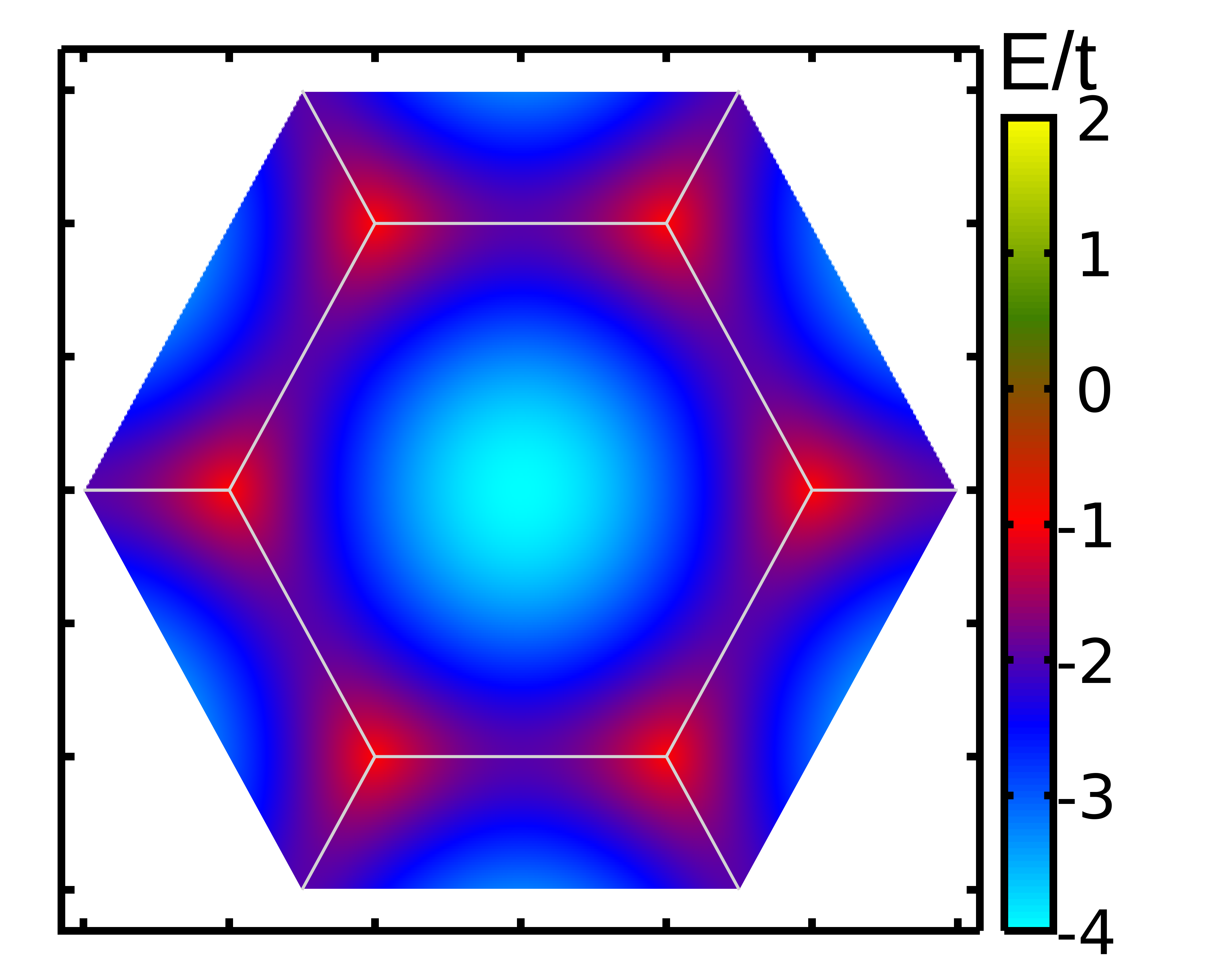}
        		%		\label{fig:kagome_lattice_delta3}
        	\end{subfigure}
        	\begin{subfigure}[b]{0.15\textwidth}
        		%   \centering
        		\subcaption{}
        		\includegraphics[width=2.8cm,height=2.2cm]{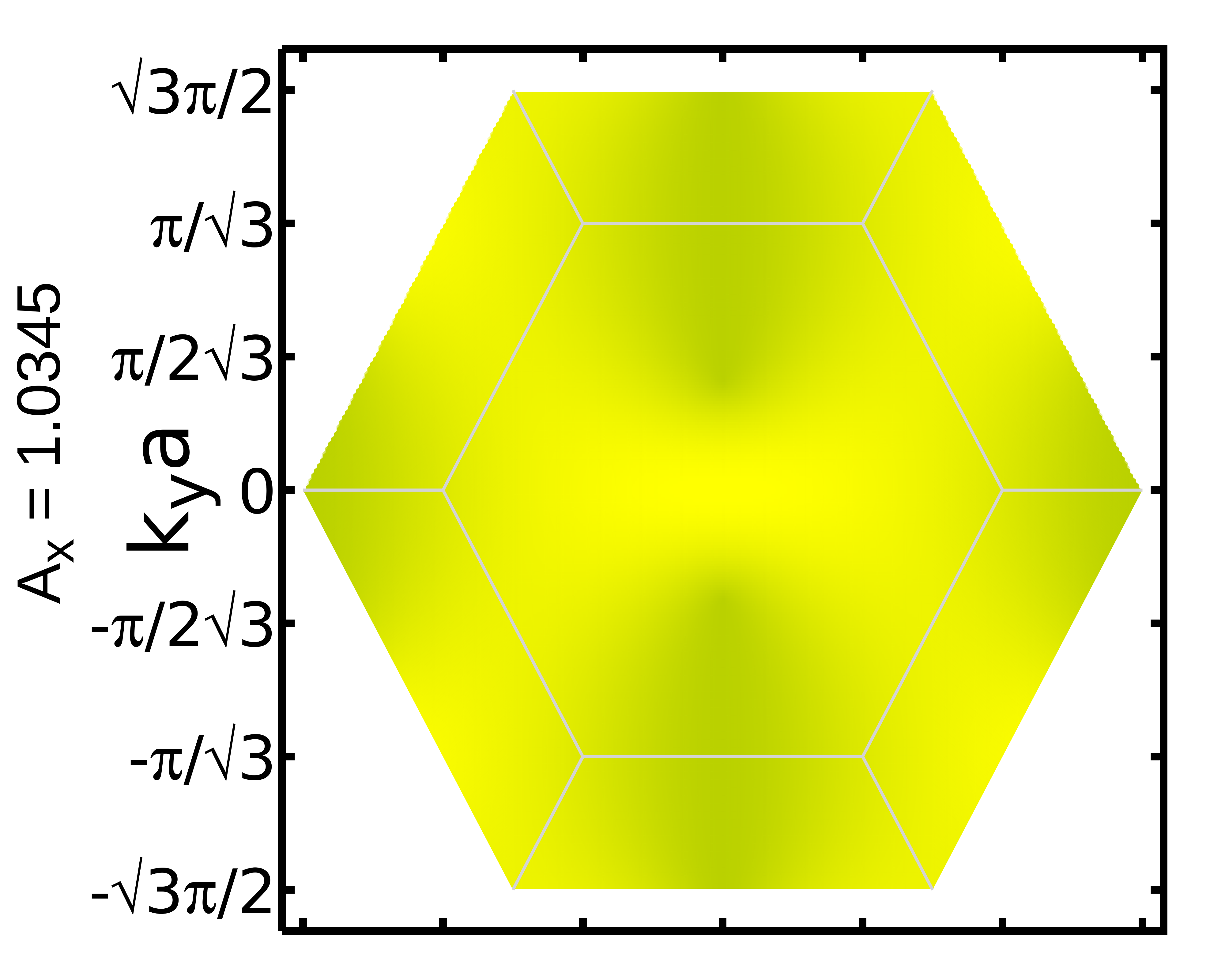}
        		%		\label{fig:GMM1M2G}
        	\end{subfigure}
        	\begin{subfigure}[b]{0.15\textwidth}
        		%   \centering
        		\subcaption{}
        		\includegraphics[width=2.5cm,height=2.2cm]{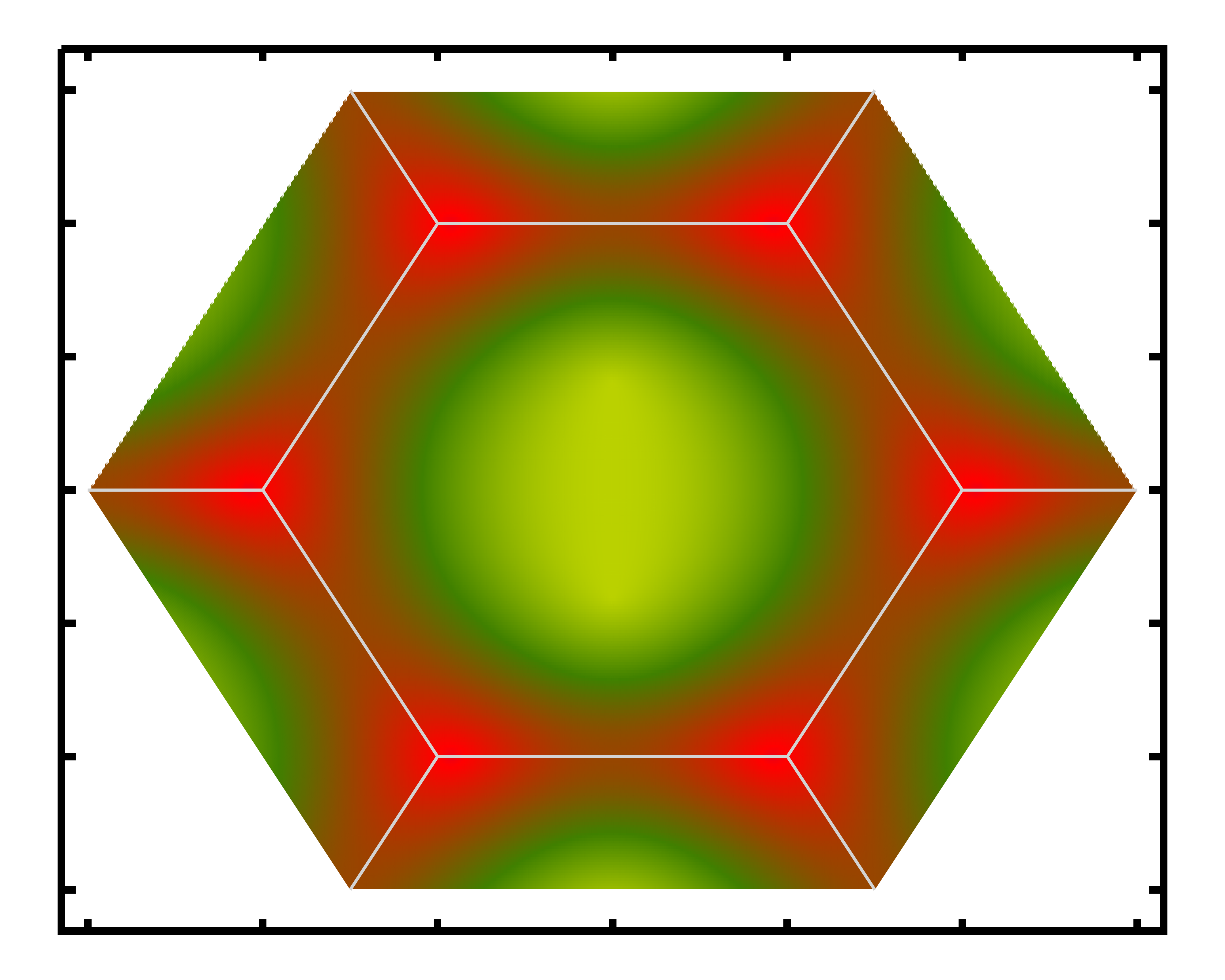}
        		%		\label{fig:kagome_delta1}
        	\end{subfigure}
        	\begin{subfigure}[b]{0.15\textwidth}
        		%   \centering
        		\subcaption{}
        		\includegraphics[width=2.8cm,height=2.2cm]{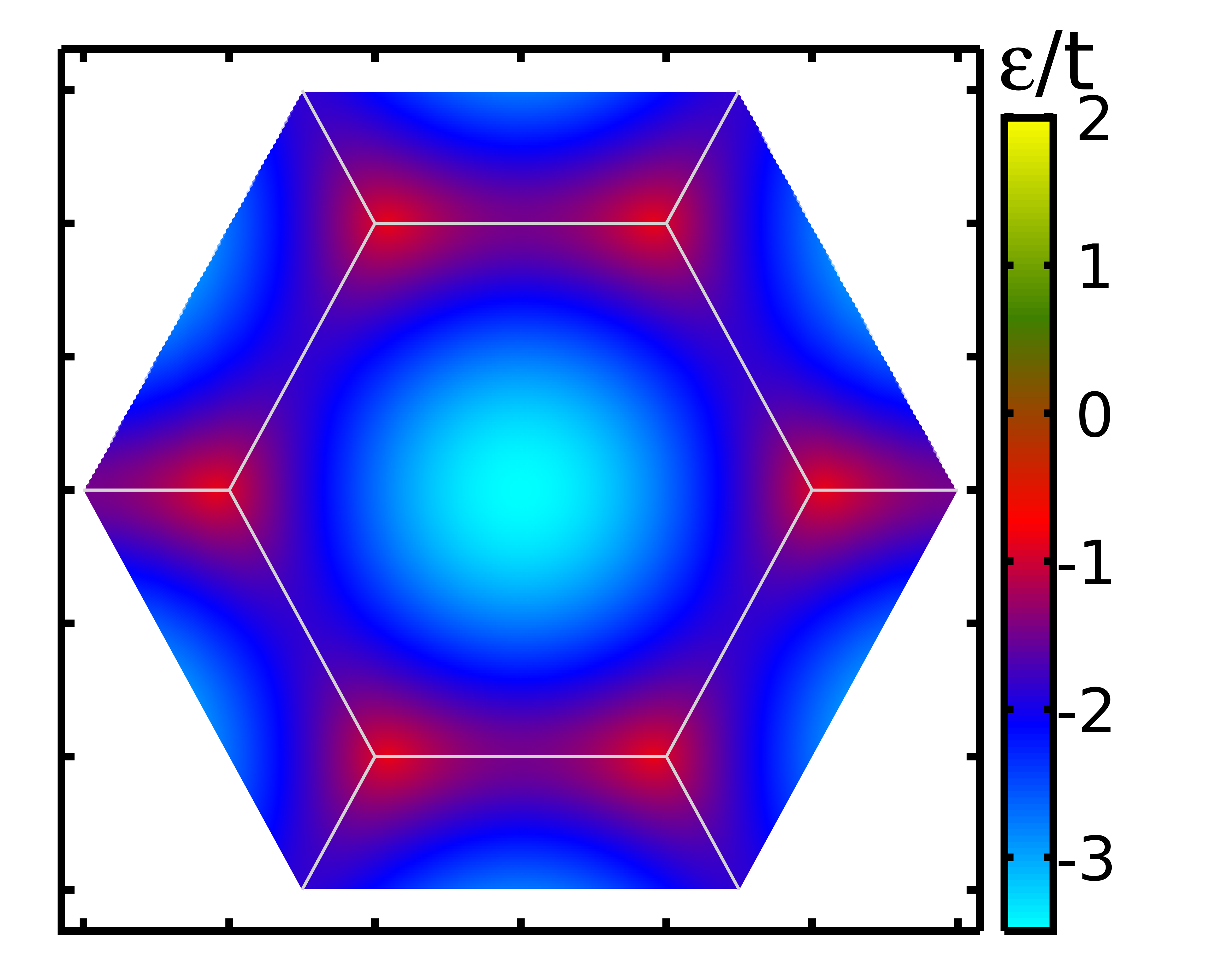}
        		%		\label{fig:kagome_delta2}
        	\end{subfigure}
        	\begin{subfigure}[b]{0.15\textwidth}
        		%   \centering
        		\subcaption{}
        		\includegraphics[width=2.8cm,height=2.2cm]{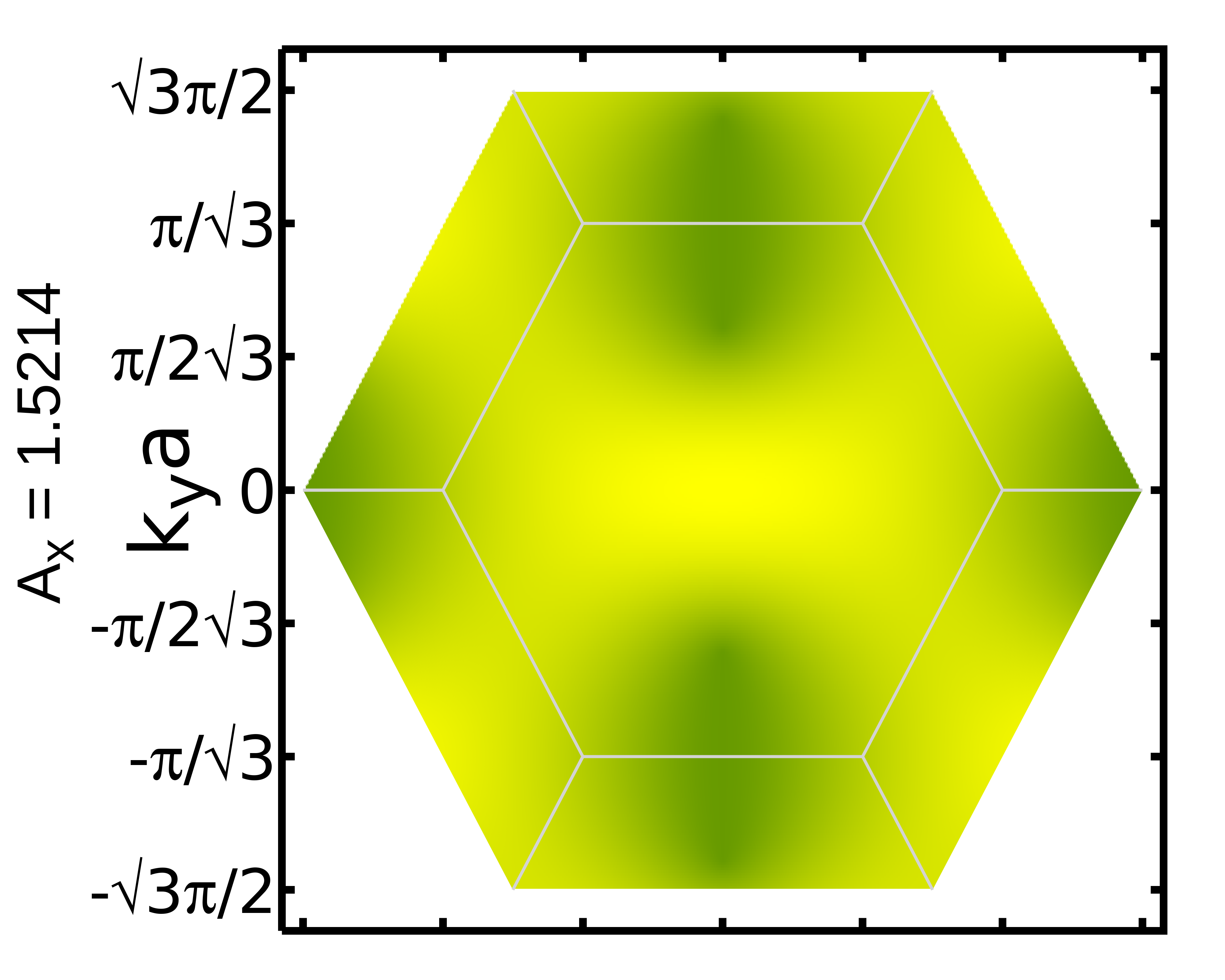}
        		%		\label{fig:kagome_delta3}
        	\end{subfigure}
        	\begin{subfigure}[b]{0.15\textwidth}
        		%   \centering
        		\subcaption{}
        		\includegraphics[width=2.5cm,height=2.2cm]{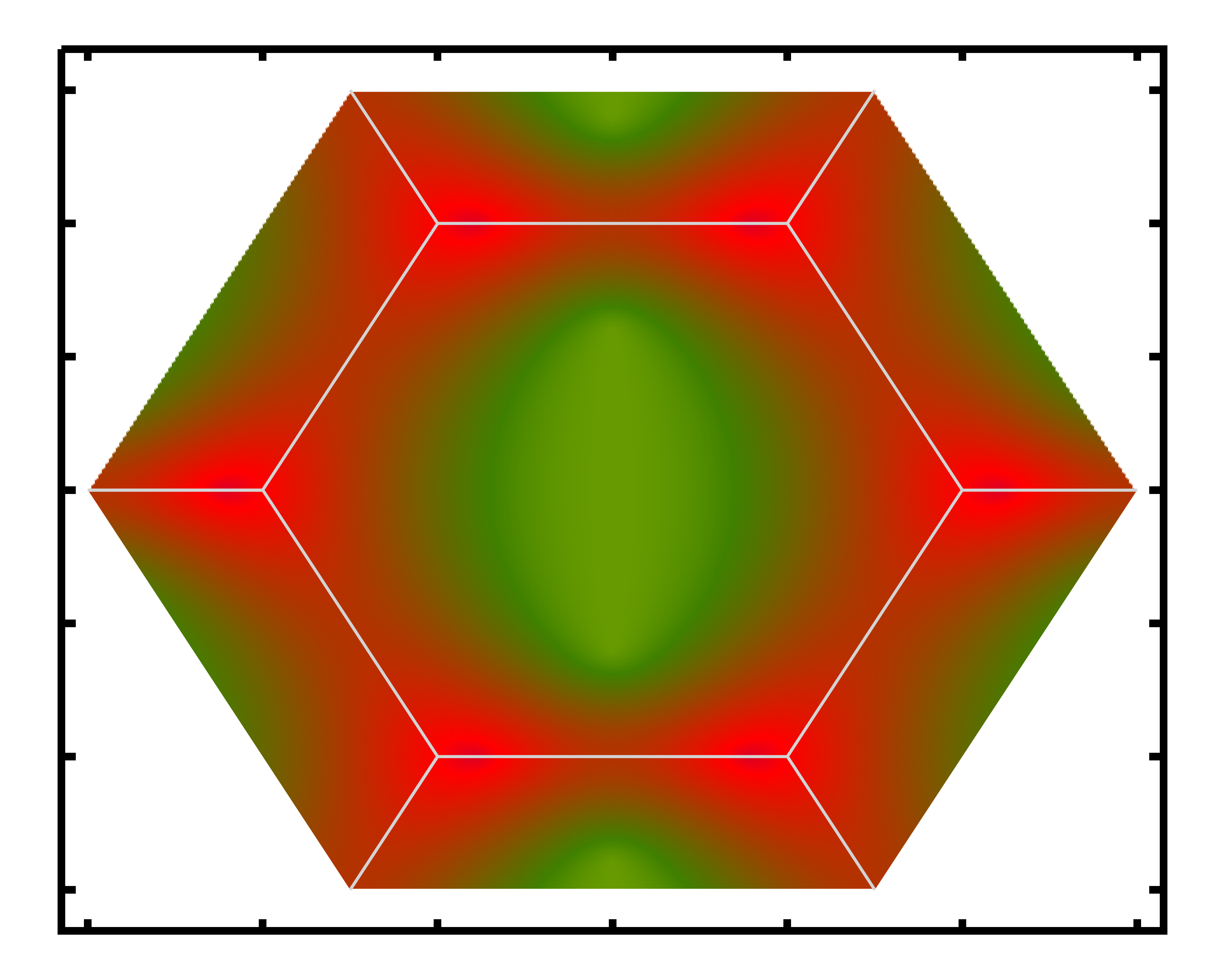}
        		%		\label{fig:liebe_67}
        	\end{subfigure}
        	\begin{subfigure}[b]{0.15\textwidth}
        		%   \centering
        		\subcaption{}
        		\includegraphics[width=2.8cm,height=2.2cm]{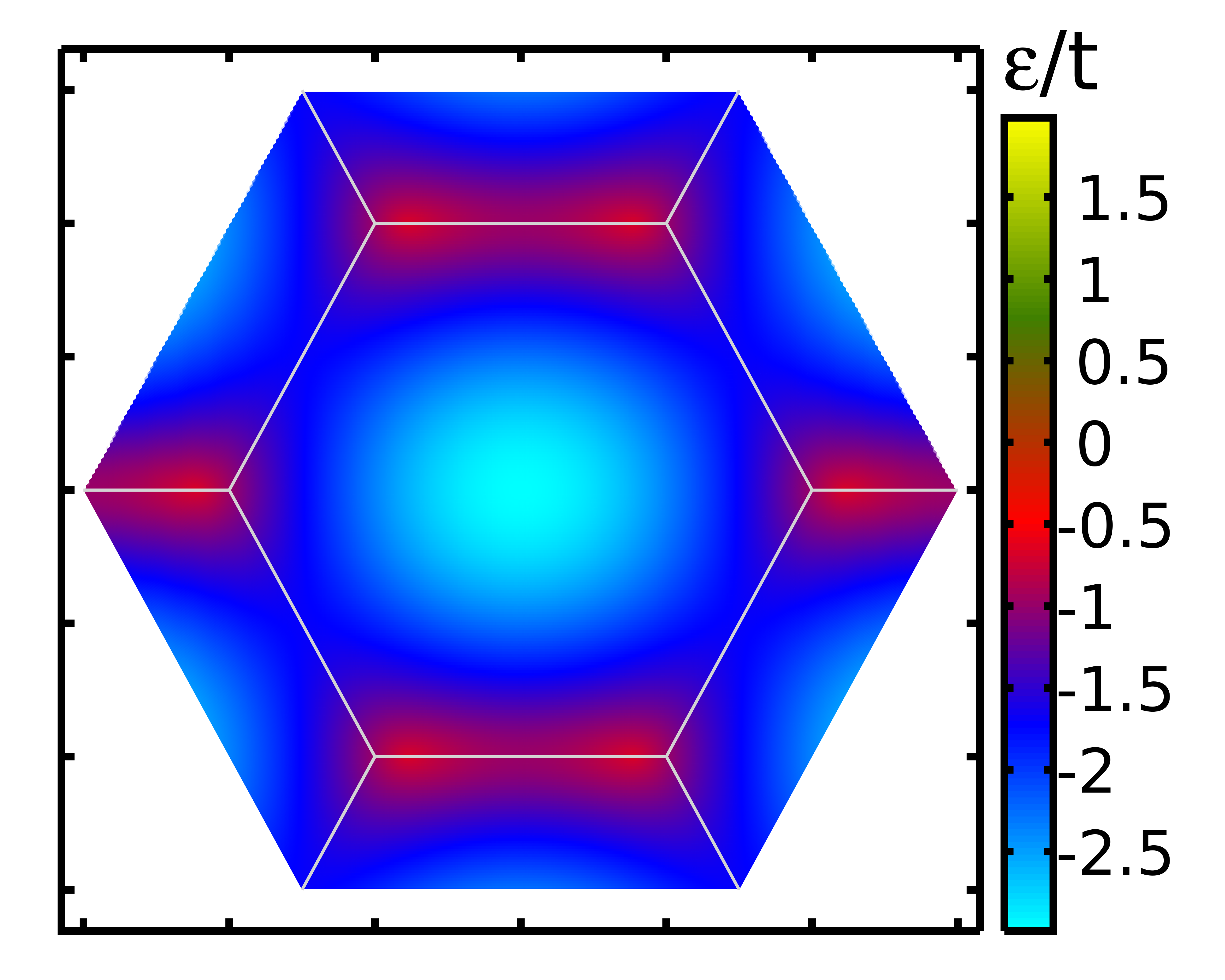}
        		%		\label{fig:kagome_edd_delta1}
        	\end{subfigure}
        	\begin{subfigure}[b]{0.15\textwidth}
        		%   \centering
        		\subcaption{}
        		\includegraphics[width=2.8cm,height=2.2cm]{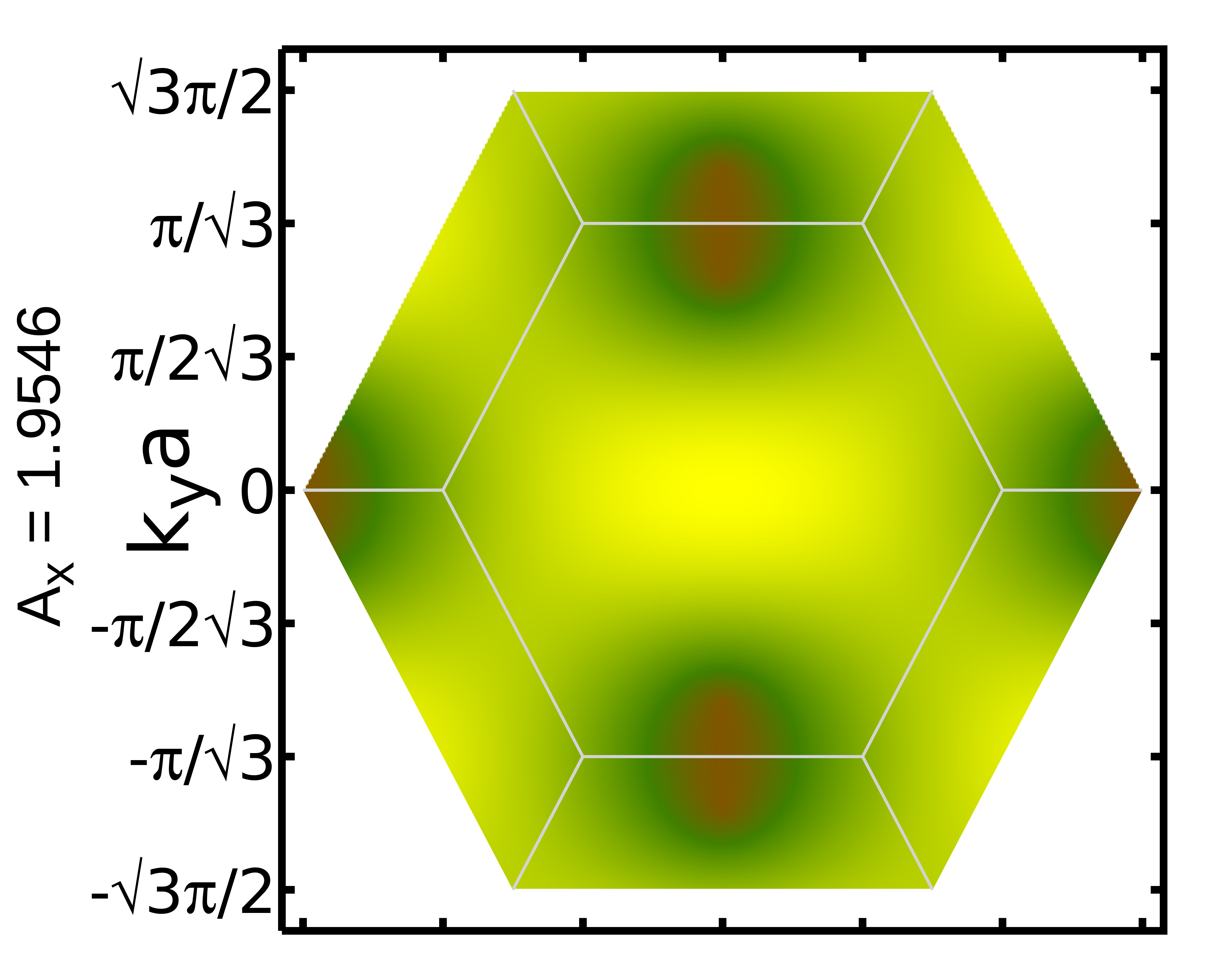}
        		%		\label{fig:kagome_edd_delta2}
        	\end{subfigure}
        	\begin{subfigure}[b]{0.15\textwidth}
        		%   \centering
        		\subcaption{}
        		\includegraphics[width=2.5cm,height=2.2cm]{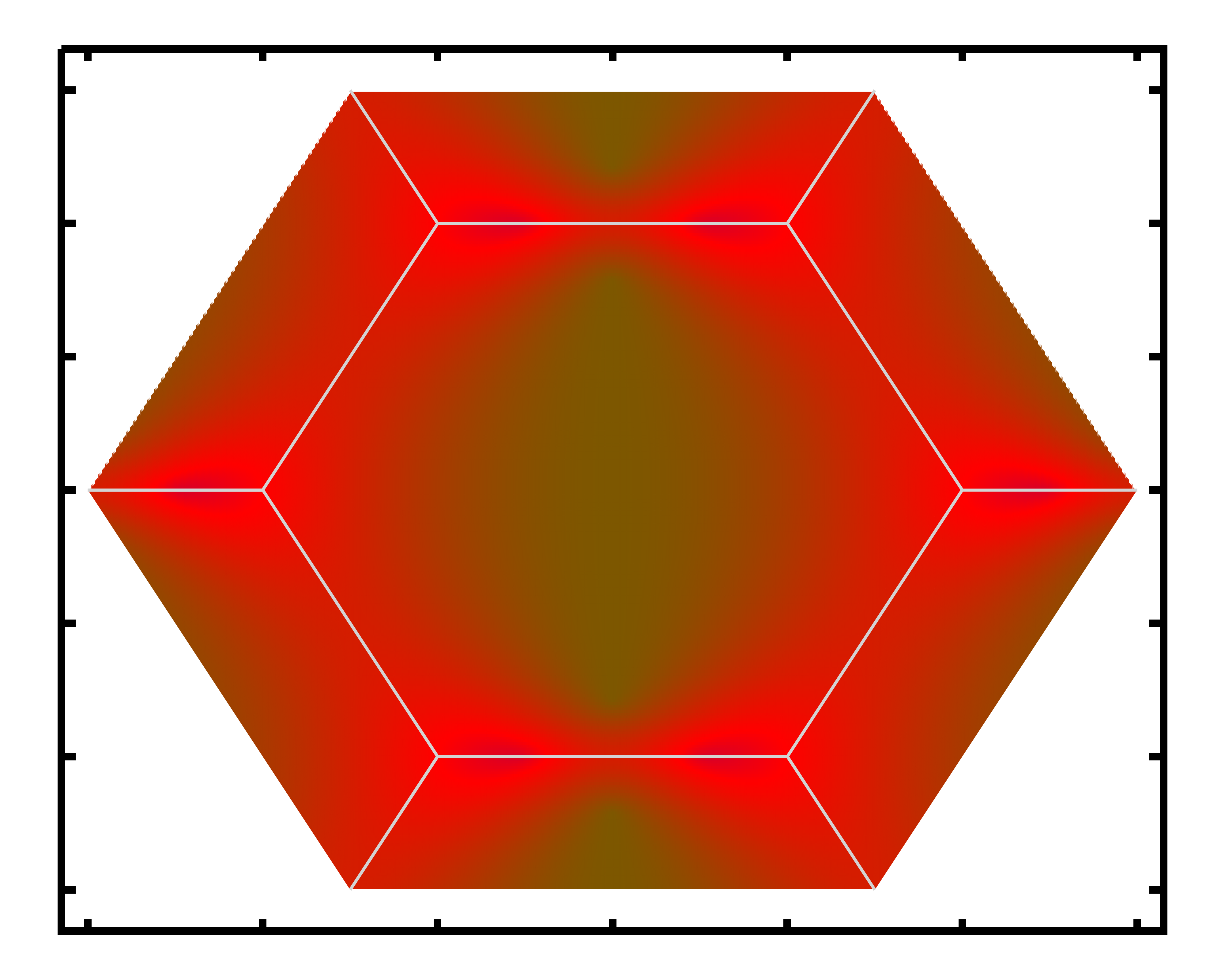}
        		%		\label{fig:kagome_edd_delta3}
        	\end{subfigure}
        	\begin{subfigure}[b]{0.15\textwidth}
        		%   \centering
        		\subcaption{}
        		\includegraphics[width=2.8cm,height=2.2cm]{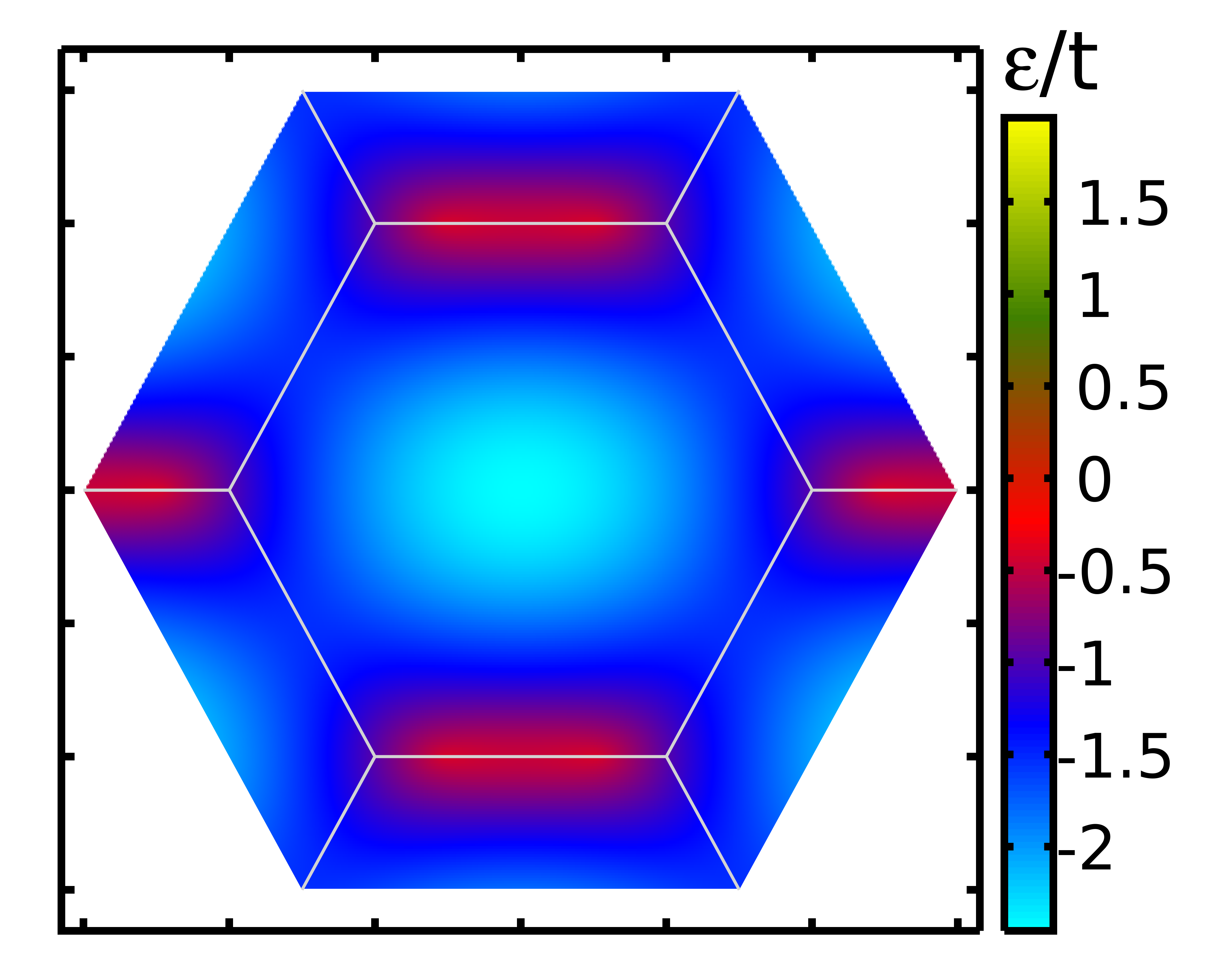}
        		%		\label{fig:lieb_edd_67}
        	\end{subfigure}
        	\begin{subfigure}[b]{0.15\textwidth}
        		%   \centering
        		\subcaption{}
        		\includegraphics[width=2.8cm,height=2.2cm]{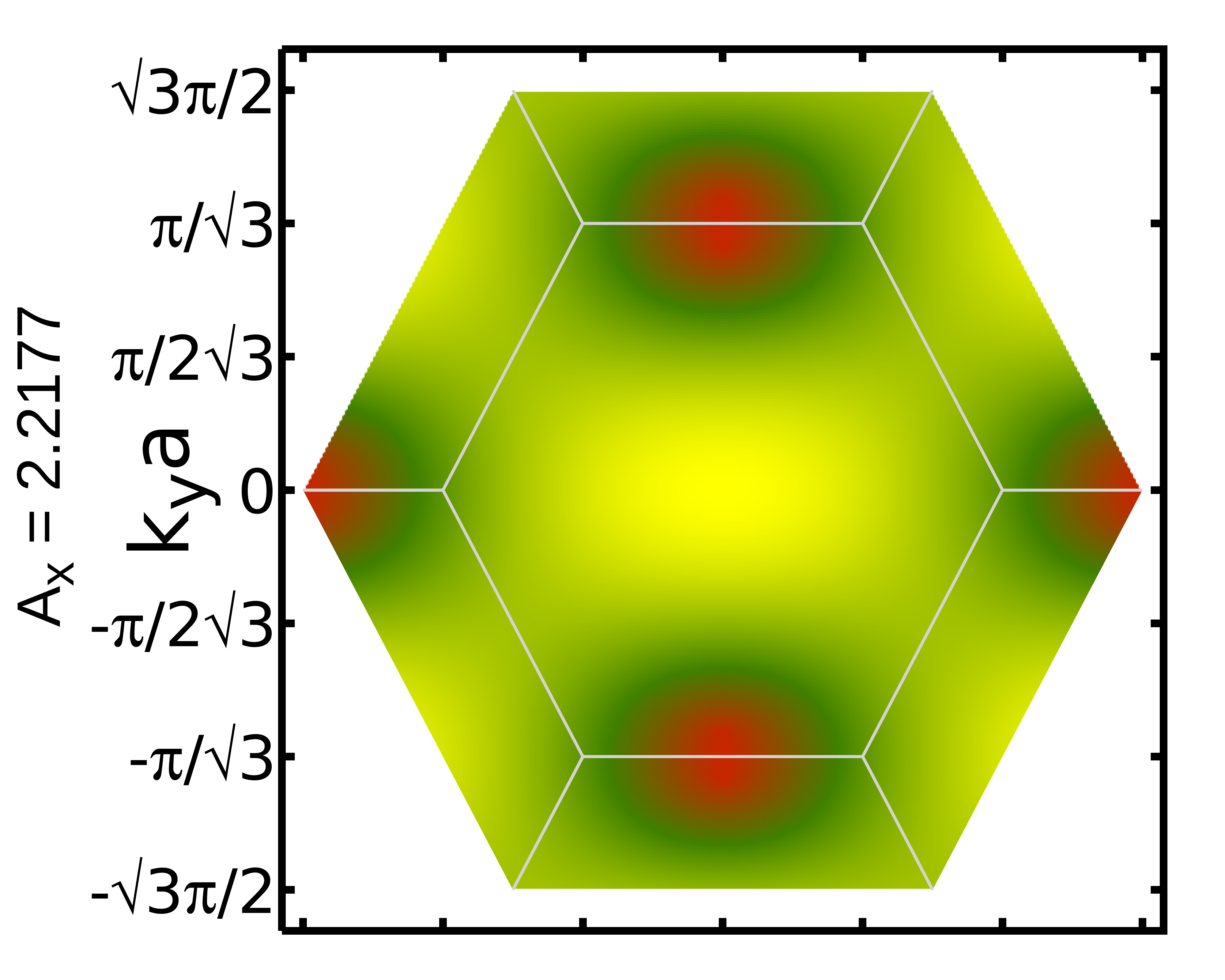}
        		%	    \label{fig:kagome_edd_delta2}
        	\end{subfigure}
        	\begin{subfigure}[b]{0.15\textwidth}
        		%   \centering
        		\subcaption{}
        		\includegraphics[width=2.5cm,height=2.2cm]{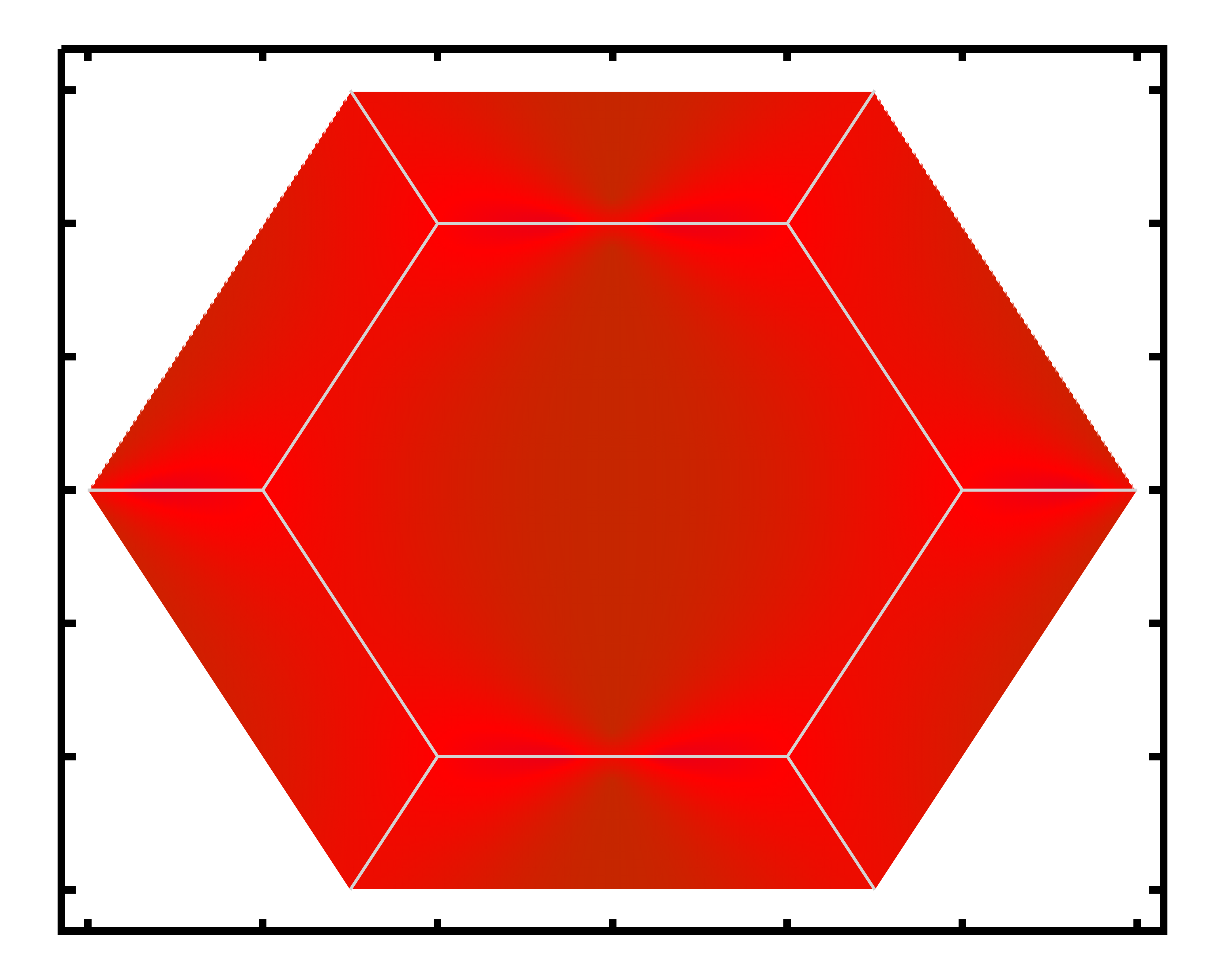}
        		%	   \label{fig:kagome_edd_delta3}
        	\end{subfigure}
        	\begin{subfigure}[b]{0.15\textwidth}
        		%   \centering
        		\subcaption{}
        		\includegraphics[width=2.8cm,height=2.2cm]{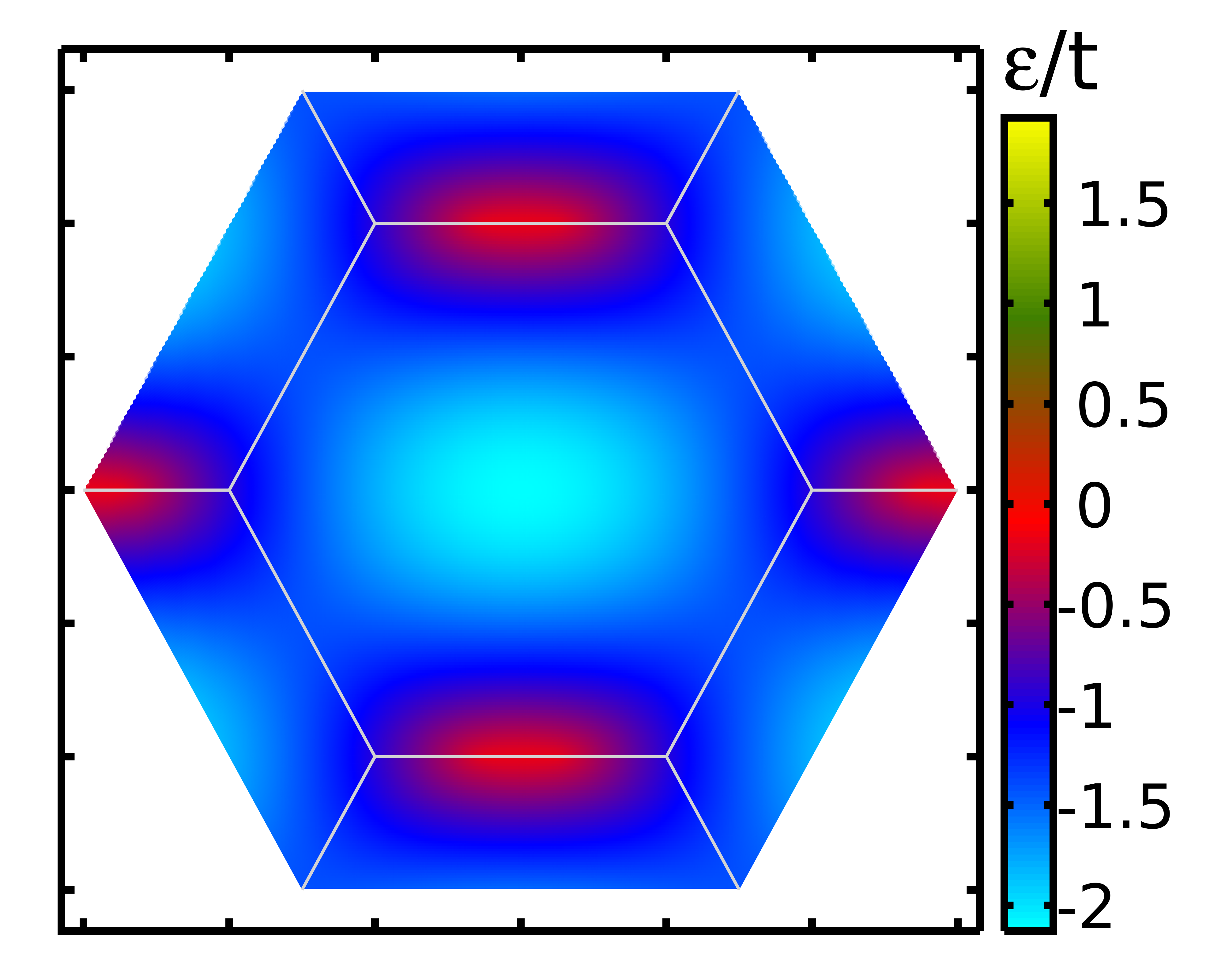}
        		%	    \label{fig:contour_lower_delta2_Ax2o2177.png}
        	\end{subfigure}  
        	\begin{subfigure}[b]{0.15\textwidth}
        		%   \centering
        		\subcaption{}
        		\includegraphics[width=2.8cm,height=2.5cm]{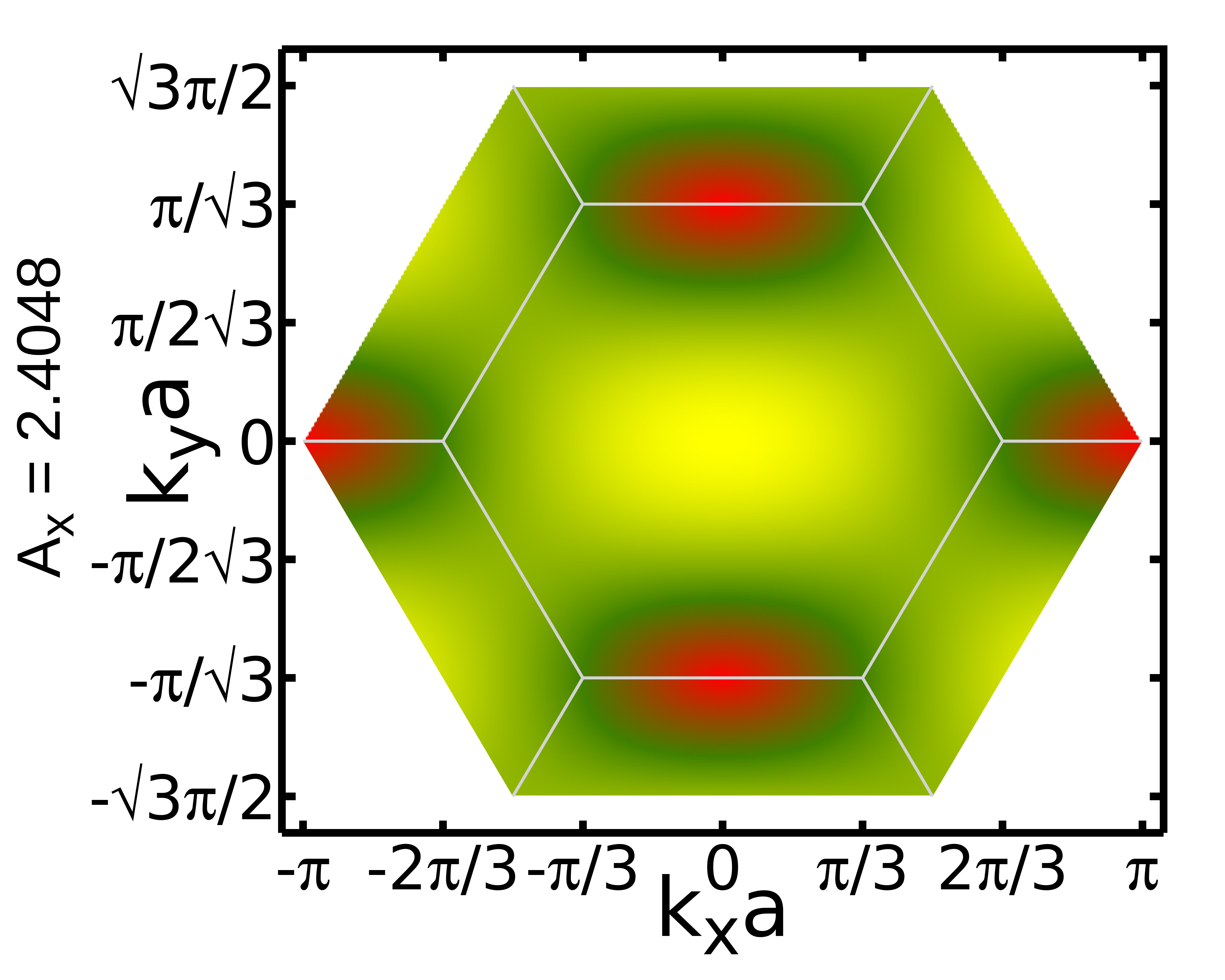}
        		%		\label{fig:contour_upper_delta2_Ax2o4048}
        	\end{subfigure}
        	\begin{subfigure}[b]{0.15\textwidth}
        		%   \centering
        		\subcaption{}
        		\includegraphics[width=2.5cm,height=2.5cm]{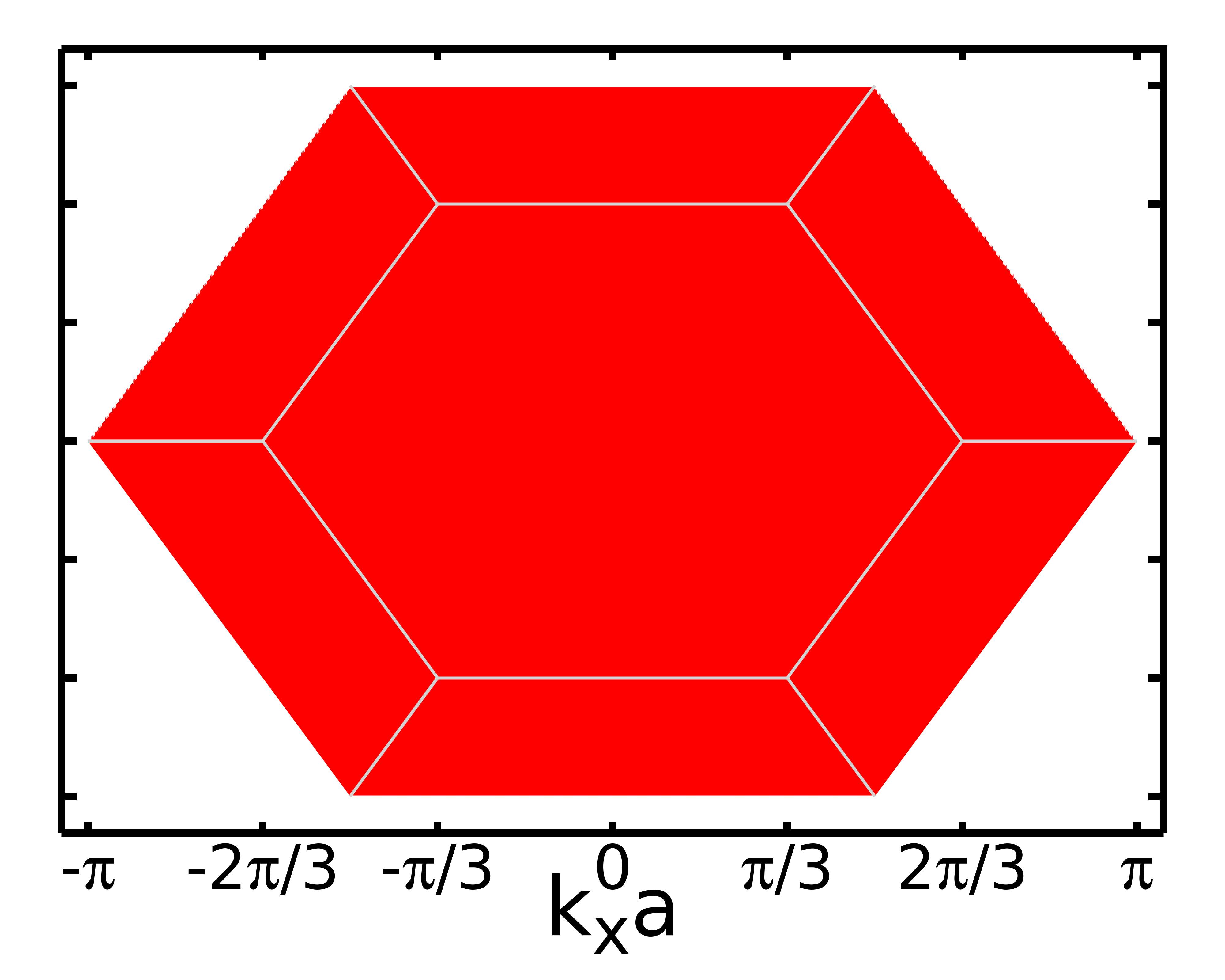}
        		%		\label{fig:contour_middle_delta2_Ax2o4048}
        	\end{subfigure}
        	\begin{subfigure}[b]{0.15\textwidth}
        		%   \centering
        		\subcaption{}
        		\includegraphics[width=2.8cm,height=2.5cm]{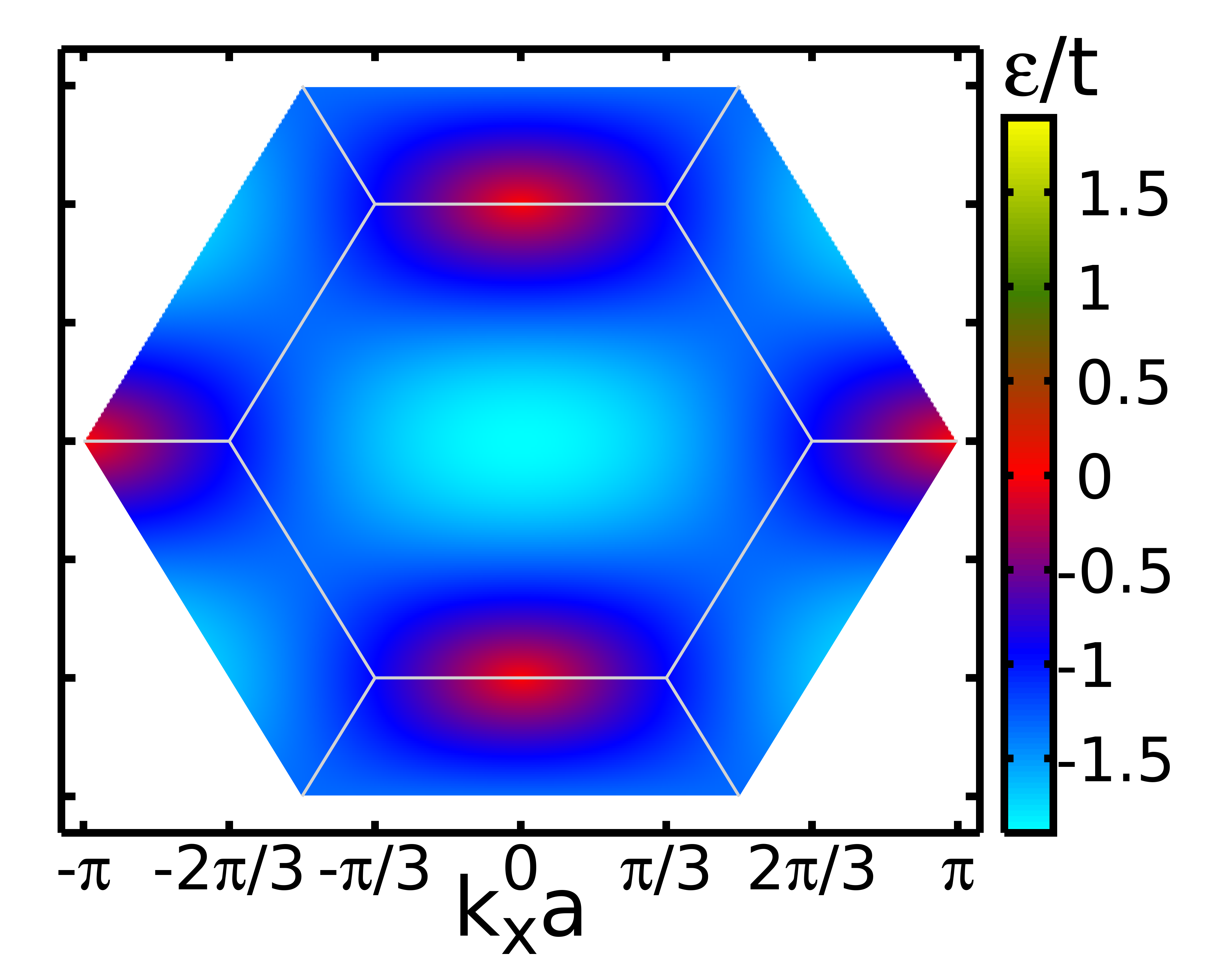}
        		%	\label{fig:contour_lower_delta2_Ax2o4048}
        	\end{subfigure}  	
        	\caption{\label{fig:kagome_merging_delta2} \justifying 2D quasienergy band structures of the kagome lattice under LPL are shown for varying $A_x$ ($A_y = 0$, $\phi = 0$). Left, middle, and right columns represent the upper, middle, and lower bands, respectively. The top panels show undressed bands, with fixed energy-level color bars displayed alongside the last column.}
        \end{figure}

		We begin our discussion by exploring the electronic structure of undriven kagome and Lieb lattices, illustrated in Fig. \ref{fig:kagome_lieb_without_light}(a) and Fig. \ref{fig:kagome_lieb_without_light}(e). Both lattices consist of three atoms per unit cell, represented as red (R), black (B), and green (G), arranged with two atoms at edge centers and one at the corner. In the kagome lattice, each atom has four nearest neighbors (NNs), defined by three nearest neighbor vectors (NNVs): $\vb{\delta}_1 = (a/2, \sqrt{3}a/2)$, $\vb{\delta}_2 = (a, 0)$, and $\vb{\delta}_3 = (-a/2, \sqrt{3}a/2)$. Its LVs $\vb{a}_1 = (a, \sqrt{3}a)$ and $\vb{a}_2 = (2a, 0)$ form a rhombic unit cell. The Lieb lattice, with LVs $\vb{a}_1 = (2a, 0)$ and $\vb{a}_2 = (0, 2a)$ forming a square unit cell, has R- and G-atoms with two NNs and B-atoms with four NNs, defined by NNVs $\vb{\delta}_1 = (0, a)$ and $\vb{\delta}_2 = (a, 0)$, as illustrated in Fig. \ref{fig:kagome_lieb_without_light}(e). For numerical calculations, we set $a = 1$. The HSPs of the Brillouin zone and the tight-binding band structures of both lattices are shown in Figs. \ref{fig:kagome_lieb_without_light}(b)-(c) and \ref{fig:kagome_lieb_without_light}(f)-(g), respectively. Both lattices feature two dispersive bands and a flat band, with the kagome flat band located at the top or bottom, determined by destructive quantum interference from lattice symmetry. In the Lieb lattice, destructive interference localizes the flat band at zero energy between the dispersive bands due to chiral symmetry. Electron density distribution (EDD) plots (Figs. \ref{fig:kagome_lieb_without_light}(d) and (h)) show kagome flat bands include contributions from all atoms, while Lieb flat bands involve only R- and G-atoms due to less coordination with NNs.

       We extend our discussion to periodically driven systems and derive modified hopping integrals for the kagome lattice using Eq. (\ref{general_hopping}). The Fourier components of the FBH are constructed using these renormalized hopping terms and truncated at a two-photon cutoff ($q \leq 2$) for computational efficiency \cite{cpb,photon_cutoff}. The quasienergy spectra, obtained by diagonalizing the FBH, can be tuned via field parameters such as $A_x$, $A_y$, and $\omega$. Your sentence is grammatically correct but could benefit from improved readability. When the external field frequency far exceeds the system's bandwidth $(\omega \gg \text{BW})$, Floquet sidebands become independent, and the FBH becomes block-diagonal, comprising identical time-independent Hamiltonians offset by $n\hbar\omega$. In this regime, band structure modification occurs via virtual photon processes, allowing us to focus solely on the central band $n = 0$ while purposefully neglecting the sidebands \cite{photon_absorption,photon_absorption2}. This enables control of electron hopping by tuning light intensity and polarization while suppressing heating through Floquet prethermalization \cite{floquet_heat1,floquet_heat2,floquet_heat3}. In contrast, in the low-frequency regime, Floquet sidebands interact, enabling one-photon or multiphoton absorption, complicating electron hopping control. Thus, our study emphasizes the high-frequency regime. 		
		
		\begin{figure}[hb]
			\centering
			\begin{subfigure}[b]{0.23\textwidth}
				%   \centering
				\subcaption{}
				\includegraphics[width=4.2cm,height=3.5cm]{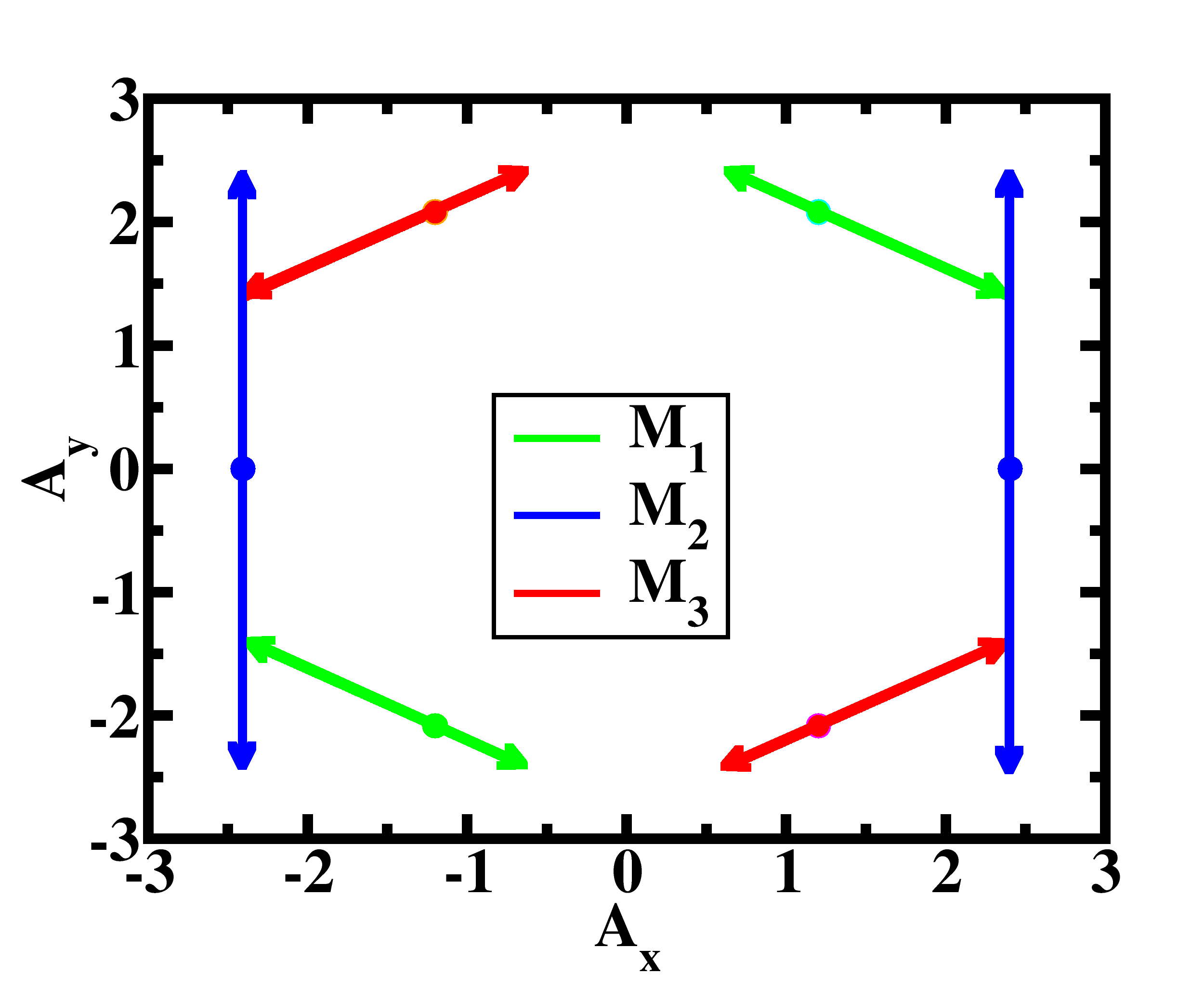}
				%		\label{fig:lieb_lattice}
			\end{subfigure}
			\begin{subfigure}[b]{0.23\textwidth}
				%   \centering
				\subcaption{}
				\includegraphics[width=4.2cm,height=3.5cm]{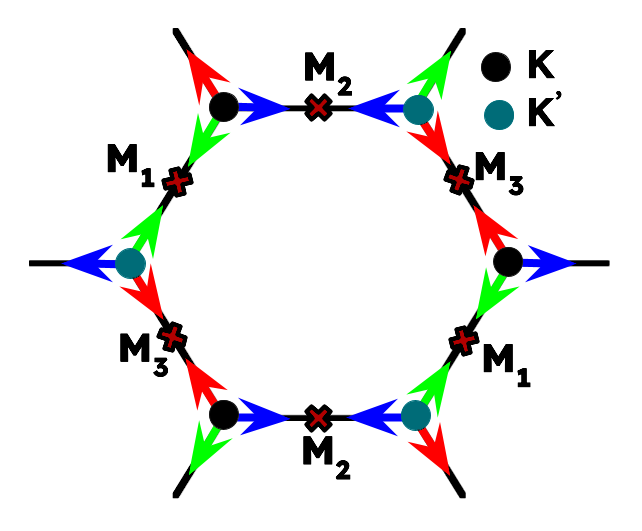}
				%		\label{fig:lieb_hsp}
			\end{subfigure}
			\caption{\label{fig:merging} \justifying (a) A merging line plot showing Dirac point merging at different HSPs for varying $A_x$ and $A_y$ with $\phi = 0$. (b) A hexagonal Brillouin zone illustrating Dirac point movement under LPL dressing.}			
		\end{figure} 
		
		In this regime, when light is applied, the Peierls phase, $e^{i \vb{A}(\tau) \cdot \delta_{j}}$, represents the phase acquired by an electron moving along the $\delta_j$. To emphasize the control of hopping strength along specific NNVs, this study focuses on LPL, where setting $\phi = 0$ for arbitrarily directed characterise by the vector potential $\vb{A}(\tau) = (A_x, A_y) \cos(\omega \tau)$. Adjusting $A_x$ and $A_y$ alters the polarization direction. Substituting these parameters in Eq. \ref{general_hopping}, we find $ t_{0,j}^F $ for the high-frequency regime along the NNVs: 
		$ t_{0,1}^F = t J_0\left(\frac{a}{2}(A_x + \sqrt{3} A_y)\right)$, $t_{0,2}^F = t J_0(A_xa)$, and $t_{0,3}^F = t J_0\left(\frac{a}{2}(\sqrt{3} A_y - A_x)\right)$.
		The hopping strength $t_{0,j}^F$, determined by the Bessel function $J_0$, exhibits oscillatory behavior, peaking when argument of $J_0$ is zero. To avoid heating and lattice damage from high-intensity fields, we limit the field amplitude to $0 \leq |A_{x/y}| \leq 2.4048$ to ensure effective control of electron hopping along specific bonds while maintaining time-reversal symmetry and preserving the lattice structure.
		
		To explore the merging of Dirac points, we choose to set LPL along $\delta_2$ bond which lies along the $x$-axis to make the calculation simple. We can deliberately set $A_y = 0$ ensuring LPL completely polarized along x-direction. The renormalised hopping strength along NNVs are expressed as: $ t_{0,1}^{F} =t_{0,3}^{F}  = t J_0\left(\frac{A_xa}{2}\right) $ and $ t_{0,2}^{F} = t J_0(A_x a)$. As $A_x$ increases, $t_{0,2}^{F}$ decreases more rapidly compared to the other two hopping strengths, since the argument of the Bessel function $J_0$ in the renormalized hopping term along $\delta_2$ is twice that of $\delta_1$ and $\delta_3$. In Fig. \ref{fig:kagome_merging_delta2}, we present 2D mappings of the quasienergy band structure for the kagome lattice with varying $A_x$. The top panels of Fig. \ref{fig:kagome_merging_delta2} show the undressed kagome lattice band structure, with Fig. \ref{fig:kagome_merging_delta2}(a) depicting the flat band and Figs. \ref{fig:kagome_merging_delta2}(b)-(c) illustrating the dispersive bands. The red regions at the corners of the hexagonal Brillouin zone (HBZ) indicate Dirac points in the dispersive bands. From second row onwards, each row displays the quasienergy bands of the kagome lattice subjected to different field amplitudes mentioned at left pannels in the plot. 		
		
		\begin{figure*}[ht]
			\centering
			\begin{subfigure}[b]{0.3\textwidth}
				%   \centering
				\subcaption{}
				\includegraphics[width=5cm,height=4cm]{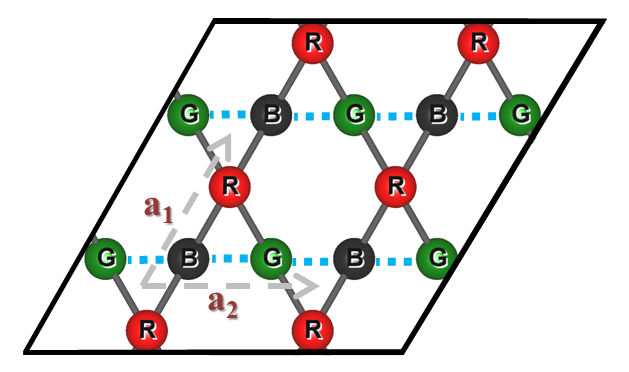}
				\label{fig:kagome_lattice_delta2}
			\end{subfigure}
			\begin{subfigure}[b]{0.3\textwidth}
				%   \centering
				\subcaption{}
				\includegraphics[width=5cm,height=4cm]{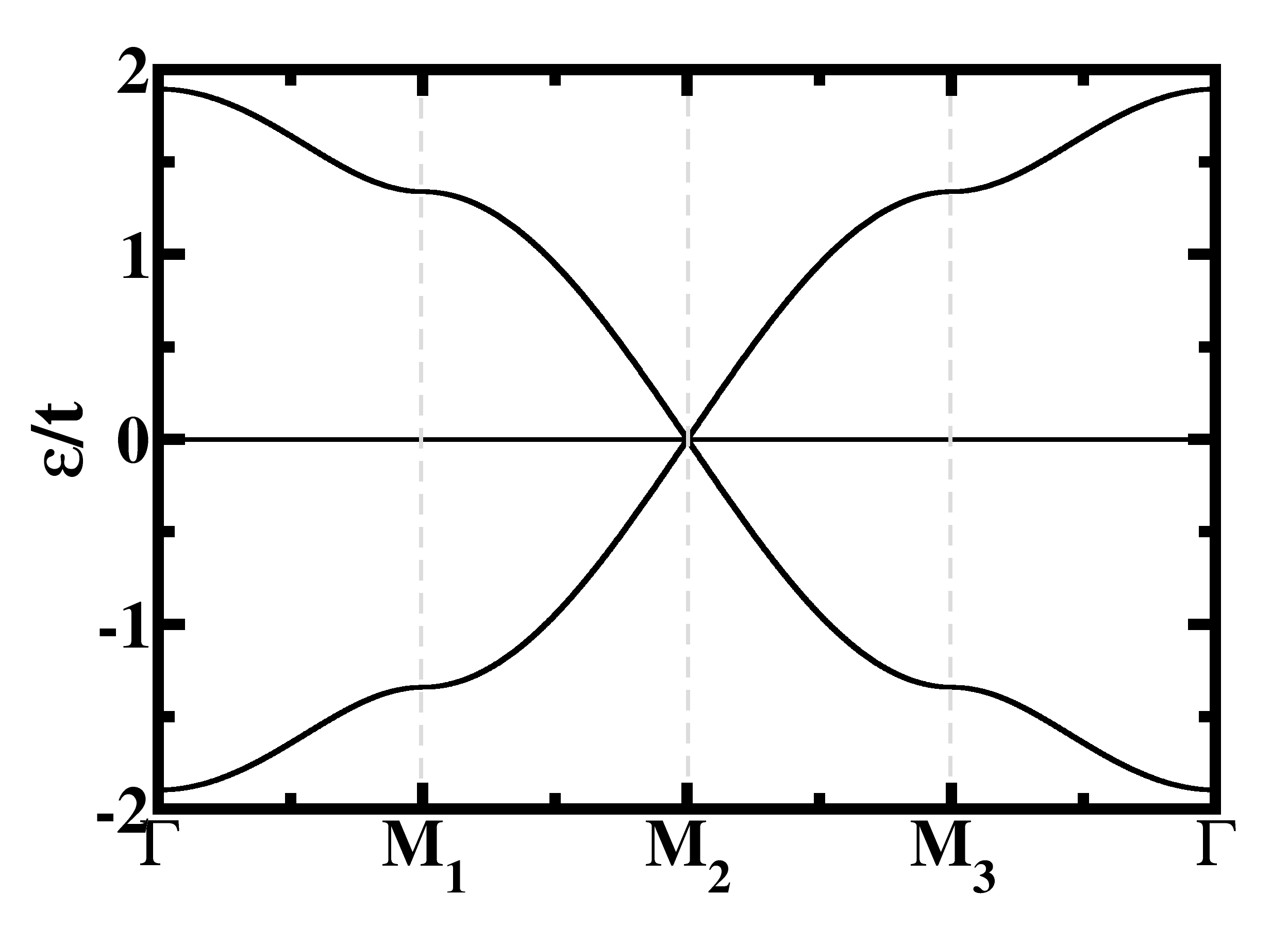}
				\label{fig:kagome_delta2}
			\end{subfigure}
			\begin{subfigure}[b]{0.3\textwidth}
				%   \centering
				\subcaption{}
				\includegraphics[width=5cm,height=4cm]{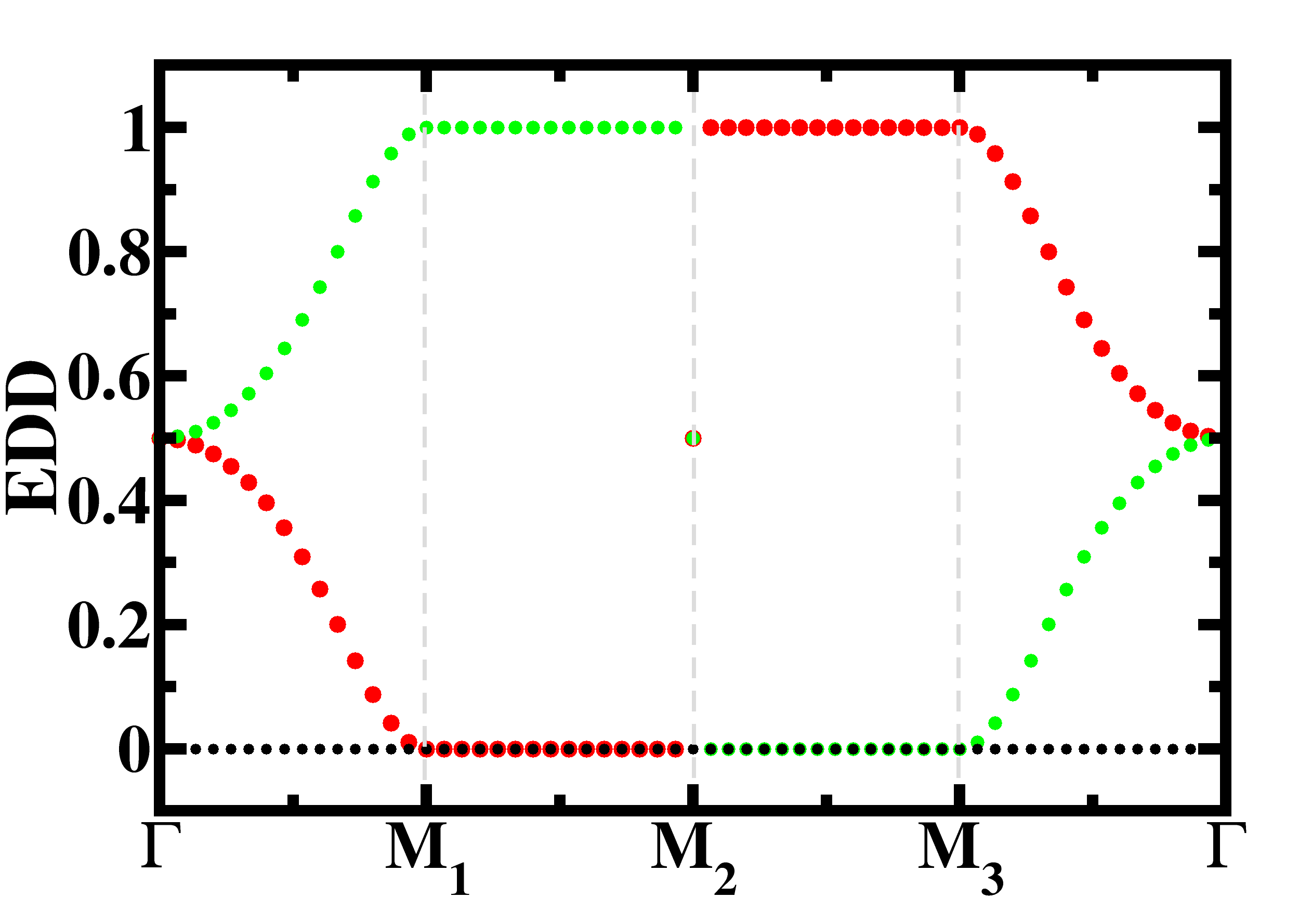}
				\label{fig:kagome_edd_delta2}
			\end{subfigure}
			\caption{\label{fig:kagome_lpl} \justifying Kagome lattices under LPL ($\phi = 0$, $A_x = 2.4048$, $A_y = 0$) in the high-frequency regime have $\delta_2$ hopping switched off (dashed line in (a)), resulting in the (b) quasienergy spectrum and (c) flat band EDD.}
		\end{figure*} 
		
		The left panels of Fig. \ref{fig:kagome_merging_delta2} clearly demonstrate that as $ A_x $ increases, the initially flat band begins to exhibit dispersive characteristics. By $ A_x = 2.4048 $, it becomes fully dispersed, with Dirac points emerging at $ \vb{M}_2 $. Simultaneously, the middle dispersive band (shown in the middle panel) becomes increasingly localized within a small region as the amplitude increases, eventually becoming completely flat. To clarify the structure of the Dirac cone, we present a 2D quasienergy band structure plot with varying amplitudes along the HSPs in Sec. 3 in SM. The transformation between the top and middle bands occurs at $A_x = 2.4048$, where hopping along $\delta_2$ is effectively suppressed $(t_{0,2}^{F} \approx 0)$, simplifying the quasienergy bands to $\varepsilon_{t/b}^{2} = \pm 2\sqrt{\sum_{i \ne 2}(t_{0,i}^{F}\cos({\vb k}\cdot{\vb \delta}_{i}))^{2}}$ and $\varepsilon_{m}^{2} = 0$, as derived from Eq. S7 (SM). Here, $\varepsilon_{t/m/b}^{2}$ denotes the top, middle, and bottom quasienergy bands of the kagome lattice and index $i$ depicts the NNVs. It is evident that turning off electron hopping transforms the middle band into a flat band at zero energy, which was previously a dispersive band in the undressed kagome structure, while the top band becomes dispersive.
			
		Our primary focus is on the right column of Fig. \ref{fig:kagome_merging_delta2}, which shows the evolution of the lower dispersive band, revealing intriguing results. As the value of $A_x$ increases, this band undergoes a transformation: the Dirac points at $K$ and $K'$ gradually shift away from their original locations toward the ${\vb M}_2$ points, as shown in Figs. \ref{fig:kagome_merging_delta2}(f) to \ref{fig:kagome_merging_delta2}(o) (right panels only). At $A_x = 2.4048$, the renormalized hopping strengths adjust to $t_{0,1}^{F} = t_{0,3}^{F} \approx 0.67t$ and $t_{0,2}^{F} \approx 0$, resulting in the complete merging of the two Dirac points at the ${\vb M}_2$ points, as depicted in Fig. \ref{fig:kagome_merging_delta2}(r). The shifting positions of the Dirac points, a direct consequence of induced anisotropy in the hopping strength along the NNs, result in the two Dirac points merging at ${\vb M}_2$. This merging phenomenon can also be described mathematically. The quasienergies at the various $\mathbf{M}_{l=1,2,3}$ HSPs can be derived from Eq. S7 (SM) and are given by: $\varepsilon_{t}^{l} = t_{0,l}^{F}$, $\varepsilon_{b}^{l} = -t_{0,l}^{F}$, and $\varepsilon_{m}^{l} = 0$. As $t_{0,2}^{F}$ decreases with increasing $A_x$, it directly affects the quasienergy states at $\mathbf{M}_2$, causing the top and bottom bands to shift symmetrically toward zero energy at the $\mathbf{M}_2$ points, as shown in Fig. S1. At $A_x = 2.4048$, this results in $\varepsilon_{t/m/b}^{2} = 0$, enforcing degeneracy at $\mathbf{M}_2$. This degeneracy results in the merging of Dirac points at $\mathbf{M}_2$ when hopping is switch off along $\delta_2$ NNVs. 
		
		We have discussed the merging of Dirac points at a specific amplitude for the simplest interaction case. Section 4 of the SM explores the effects of switching off hopping along other NNVs for various field amplitude combinations that lead to Dirac point merging, helping us identify the merging line in the $A_x$-$A_y$ plane (Fig. \ref{fig:merging}(a)), where Dirac points converge at the HSPs (Fig. \ref{fig:merging}(b)). The intersection of the two merging lines is also discussed in Sec. 5 of the SM, following the analysis of the merging line in the $A_x$-$A_y$ plane. This shows that switching off electron hopping along specific NNVs can merge Dirac points at HSPs, a challenging effect to achieve with uniaxial strain alone, as noted by Xu $et \, al. $ \cite{strain_kagome2}.
		
		An interesting aspect of the evolution of the band structure and the merging of the Dirac points at HSPs in the kagome lattice is that it results in a transition from the kagome quasienergy band structure to the Lieb band spectrum, with a reduced bandwidth. However, for the evolved quasienergy spectrum to behave like a Lieb lattice band structure, the other two renormalized hopping strengths must be equal, as they are in the Lieb lattice. Therefore, we choose the amplitudes in such a way that they lie on the merging line in Fig. \ref{fig:merging}(a) and satisfy the Lieb lattice condition.

		When electron hopping along $\delta_2$ is turned off, the values of $A_x$ and $A_y$ must lie on the blue merging line. The conditions where $t_{0,2}^{F} \approx 0$ and $t_{0,1}^{F} = t_{0,3}^{F}$ occur at $(A_x, A_y) = (2.4048, 0)$ and $(-2.4048, 0)$, represented by blue dots in Fig. \ref{fig:merging}(a). This configuration causes the Dirac points to merge at $\vb{M}_2$ (Fig. \ref{fig:merging}(b)), with hopping strengths along $\delta_1$ and $\delta_3$ reduced to approximately $67\%$ of their undressed values, restricting electron movement to two directions with equal strength: along $\delta_1$ (from R-atoms to B-atoms) and $\delta_3$ (from R-atoms to G-atoms).
		
		Consequently, R-atoms function as corner atoms, while B- and G-atoms serve as edge-center atoms of the unit cell. In Fig. \ref{fig:kagome_lpl}(a), the dashed lines show the directions along which the electron hopping is turned off, effectively limiting interactions between B-atoms and G-atoms. Hence, R-atoms have coordination number $4$ and B- and G-atoms have coordination number $2$ like the Lieb lattice. Under this amplitude configuration, two dispersive bands linearly meet with a flat band at zero energy, forming a Dirac cone, as shown in Fig. \ref{fig:kagome_lpl}(b). The EDD plots of the flat band, which appears at zero energy is shown in Fig. \ref{fig:kagome_lpl}(c). From this plot, it is clear that the flat bands at zero energy are predominantly contributed by atoms that act as edge centre atoms in the unit cell. By comparing Fig. \ref{fig:kagome_lpl}(b) and Fig. \ref{fig:kagome_lieb_without_light}(g), we can conclude that the quasienergy spectrum of the driven kagome lattice qualitatively resembles the band structure of the Lieb lattice. Again, comparing EDD of flat band in driven kagome lattice (Fig. \ref{fig:kagome_lpl}(c)) and undriven Lieb lattice (\ref{fig:kagome_lieb_without_light}(h)) verify that in both cases flat band have contribution from only edge-centres atoms. Secction 6 of the SM presents other amplitude values where the quasienergy spectrum of the kagome lattice resembles the Lieb lattice band structure. Thus, we can delebrately state that merging of Dirac points at different HSPs in kagome lattice under LPLs at specific value of amplitude can leads to transition from kagome spectrum to Lieb like spectrum.

       In this work, we derive a generalized expression for renormalized hopping strength in 2D lattices under periodic driving fields, applicable to any system with arbitrary polarizations, as presented in Eq. (\ref{general_hopping}). In the high-frequency regime, LPL with varying field parameters allows precise control over the hopping strength along NNVs. This control enables significant modifications to the lattice band structure, facilitating the tuning of various physical properties such as longitudinal and Hall optical conductivities \cite{optical1,optical2,optical3}, transport \cite{transport1,transport2}, tunneling conductance \cite{tunneling1,tunneling2}, and other system-specific behaviors. For specific light amplitudes, the hopping strength along chosen NNVs can be completely suppressed which leads to the merging of two Dirac points related by time-reversal symmetric. When hopping is turned off along the $\delta_i$ direction, the Dirac cones merge at the corresponding $\vb{M}_i$ point. Furthermore, when the other two hopping strengths equalize for particular values of $A_x$ and $A_y$, the band structure of the kagome lattice transitions into the Lieb lattice spectrum, with a reduced band width.
			
		\section*{Supplementary Material }
		The supplementary material includes a detailed discussion of Floquet-Bloch theory. It also presents the mathematical expressions for the quasienergy spectrum of the kagome lattice and examines its band evolution under exposure to linearly polarized light. Additionally, it covers the merging of Dirac cones at various HSPs and concludes with EDD plots. 

		P.P. acknowledges DST-SERB for the ECRA project (No. ECR/2017/003305).\\
		\section*{Conflict of interest}
		The authors declare no conflict of interest.
		\section*{Author Contributions}
		\textbf{Gulshan Kumar}: Conceptualization (lead); Data curation (equal); Formal analysis (equal); Investigation (equal); Methodology (equal); Writing – original draft (lead); Writing – review and editing (equal). 
		\textbf{Shashikant Kumar}: Conceptualization (equal); Formal analysis (equal); Visualization (equal); Writing – review and editing (equal). 
		\textbf{Prakash Parida}:   Conceptualization (equal); Supervision (equal); Validation (equal); Visualization (equal); Writing – review and editing (equal).
		
		\section*{DATA AVAILABILITY}
		The data that support the findings of this study are available from the corresponding author upon reasonable request.

		\bibliography{main.bib}
		
	\end{document}